\documentclass[times]{kristall}	        

\usepackage{epsf}

\begin{document}
\maketitle		

\begin{abstract}	
\Abstract		
\end{abstract}		


\newcommand{\onlinecite}{\cite}
\newcommand{\rem}[1]{}
\newcommand{\VEIT}[1]{}
\newcommand{\MAREK}[1]{}

\def \AA {{{\rm A}{\kern -0.1cm}^{\circ} }} 
\def \AC {{\cal A}} 
\def \pe {^\perp}
\def \h {{{\bf h}^\perp}}  
\def \hperp {{{\bf h}^\perp}}  
\def \uperp {{{\bf u}^\perp}}  
\def \uu {{\bf u}}  
\def \r {{\bf r}}  
\def \rhat {{\rho^{\rm atom}}} 
\def \rhoo {{\rho_0}} 
\def \rhoA {{\rho_\AC}} 
\def \rhoh {{\rho_h}} 
\def \rhog {{\rho_g}} 
\def \trho {{\tilde \rho}} 
\def \trhoo {{\tilde \rhoo}} 
\def \x {{\bf x}}  
\def \rand{{\rm rand}}   
\def \id{{\rm ideal}}   

\def \GGperp {{{\bf g}^\perp}}
\def \GGpar {{\bf g}}
\def \Gpar {g}
\def \Gperp {{g^\perp}}

\def \q {{\bf q}}
\def \qq {{\bf q}}
\def \R {{\bf R}}
\def \u {{\bf u}}

\def \Ftile {{F^{\rm tile}}}
\def \Fdeco {{F^{\rm deco}}}


\section{Introduction}
\label{sec:intro}

The ``random tiling'' model of quasicrystals has 
specific and sometimes radical implications for the
procedures that should be used to determine the atomic structures.
This paper collects several kinds of answer to the 
question, ``How should one modify the standard crystallographic 
procedures in order to solve the atomic structure of a random tiling 
quasicrystal?'' 
(Notice that even the definition of 
``solve'' is arguable in the case of an intrinsically random structure.)
We intend this paper to encourage 
crystallographers to try 
random-tiling fits to diffraction data,
by showing that a couple of approaches have already been thought out 
in some detail. It also cautions against some common misunderstandings
about this point, and collects three new (and imperfectly digested)
numerical tests of our ideas.
The discussion goes well beyond its antecedent in Sec.~7 of
Ref.~\onlinecite{Hen91ART}. 


\subsection {Perp space and diffraction}
\label{sec:perpspace}

The ``perp'' coordinate $\hperp$ is a central notion in
quasicrystallography. Before proceeding, 
we had better remind the reader that it has two definitions, 
which coincide for the case of an ideal tiling.
\begin{itemize}
\item[]
{\it Definition 1, } valid for 
the usual higher-dimensional ``cut'' description~\cite{Janot}
of any quasiperiodic structure (e.g. an ideal tiling).
The (periodic) higher-dimensional
density consists of atomic surfaces extending in the perp direction;
then $\hperp$ is the perp-space displacement from the 
intersection of the cut plane and the atomic surface, to a
vertex of the higher-dimensional lattice.

\item[]
{\it Definition 2}, 
valid for any arbitrary (e.g. random)  tiling. 
Any tiling can 
be ``lifted'' so that each tile vertex maps to a vertex of 
a periodic lattice in a higher dimension~\cite{Hen91ART,Els85b};
then $\hperp$ of each tile vertex is the difference between the
perp component of its lattice vector and the mean value
(``center-of-mass'' in perp space). 
\end{itemize}

Consider the probability distribution of the Definition 2 $\hperp$ values.
It is easy to check that, as long as the physical (hyper)plane has
an irrational orientation, 
and in the limit of a large physical-space volume $V$, 
there are $n_0 V d^3 \hperp$
possible perp-space points in a perp volume element $d^3\hperp$,
where $n_0^{-1}$ is the volume of the unit cell in the
-dimensional lattice. 
The mean number of such points actually present in the tiling
is defined to be $\rho(\hperp) n_0 V d^3\hperp$, so 
$\rho(\hperp) \in [0,1]$ is a  probability.
In random tilings it is, in practice,  a smooth function. 
(In the special case of an ideal tiling,
$\rho(\h)$ is 1 within the 
``acceptance domain'' where sites are sure to be occupied, 
and zero outside it.)
Note $\hperp$, according to Definition 1, or its mean
$\overline{\hperp} \equiv \int d^3\hperp \hperp \rho(\hperp)$ 
according to Definition 2,  plays the role for 
perp displacements 
that a crystal's center-of-mass plays for ordinary displacements. 

Now, we may take an ensemble average of
a random tiling ensemble, just constraining the 
value $\overline{\hperp}$ to be zero (without loss
of generality). 
This gives, in  real space, a nonzero density 
$\rho(\r)$ on a discrete (but dense) set of points $\r$. 
It is a fact, natural though not rigorously proven, that $\rho(\r)$
depends only on the perp-coordinate $\hperp$ of the point $\r$. 
If so, it is easy to see 
$\rho(\r) =\rho(\hperp)$ as defined in the last paragraph.
In other words, starting with Definition 2, we found
the averaged density $\rho(\r)$ is given by
the cut construction of Definition 1, if the atomic
surfaces are weighted by $\rho(\hperp$. 
This shows that $\rho(\r)$ is quasiperiodic, since 
that term is, practically speaking,  {\it defined} 
by the existence of a cut construction.  
\footnote{In the case of the two-dimensional equilibrium random 
tiling or the icosahedral glass, the above arguments break
down because $\rho(\hperp) =0$, i.e. the distribution gets broader
and broader with increasing $V$.}

These facts about $\rho(\hperp)$ easily generalize 
to atomic decorations in which  atoms sit on some of the 
tile vertices.
Thus wherever the physical 3-plane cuts an atomic surface,
$\rho(\hperp)$ is the probability that an atom is present.

The Bragg component of the diffraction amplitude $F_\GGpar$ 
is defined as the part which scales as $V$ with the volume. 
It is,  in fact, simply the Fourier transform of the averaged density:
~\cite{Els85b}, 
   \begin{equation}
       F(\GGperp) = \int d^3  \h \rho(\h) e^{-i \GGperp\cdot \h} 
   \label{eq:FQFT}
   \end{equation}
Here $\GGperp$ is the perp-space
partner of the reciprocal lattice wavevector $\GGpar$.
In cases other than icosahedral, ``3'' in eq.~(\ref{eq:FQFT}) is
replaced by  ``$d_\perp$'', the dimensionality of perp space.
(However we shall assume the physical dimension is always 3, since
this paper considers determinations from Bragg peaks, but
random tilings have no Bragg peaks in dimension 1 or 2.)
For a tiling decorated by real atoms, 
(\ref{eq:FQFT}) can be suitably generalized by including
form factors and allowing for different classes of site
(see eq.~(\ref{eq:Fdeco}), below).

\subsection {Random tilings and the cut description}

Bragg diffraction amplitudes are 
always represented by a quasiperiodic density 
in higher-dimensional space, 
which when cut by a 3-plane at the correct incommensurate 
orientation gives the diffracting density in physical space.~\cite{Janot} 
The crystallographic refinements done to date 
have presupposed that the structure,  ideally, 
{\it is} quasiperiodic.
But an (at least) equally plausible scenario is that
the thermal equilibrium structure is an intrinsically 
random tiling and 
so that its long-range quasiperiodic order is
stabilized by the tiling's configurational entropy.~\cite{entropic,Hen91ART}
Notions 
concerning the free energy and elastic constants are
central to the {\it physics}~\cite{entropic,Hen91ART} but
will not be repeated here since they are somewhat tangential 
to the {\it crystallography}. Indeed, large parts of this
paper could be applied equally well to the cases of
(i) the ``icosahedral glass''~\cite{iglass}, e.g. $i$(AlCuLi), or
$i$(TiNiZr), which is not thermodynamically stable;
or (ii) the ``weak matching-rule'' quasicrystal~\cite{weakrules}, 
which is a slight generalization of the ideal, energetically
stabilized quasicrystal that nevertheless has a nonzero density
of intrinsically random sites.

So, how does the cut description get modified in the case
of a random tiling?
As just noted in Sec.~\ref{sec:perpspace}, 
the physical-space density giving rise to the Bragg amplitudes is the 
ensemble average of the scattering density. 
This average is perfectly quasiperiodic, and so even for
a random tiling it is represented as a cut
by a 3-plane through a function in a higher-dimensional space.  

However, the physical-space density will have a great many
fractionally occupied atoms; even if there is no substitutional 
disorder, since a given volume of space has probabilities to
be divided into tiles in different ways and different tiles
have different decorations. 
A second effect is the 
``physical-space displacements''~\cite{Bois90}, 
the deviations of an atom from ideal tiling vertex
positions in response to forces from neighboring atoms.
In ideal structures, an atom's local environment is a
deterministic function of $\hperp$. Then the atom's equilibrium position
is $\r_0 + \uu(\hperp)$, also a function of $\hperp$, and this
is the 
parametrization of the atomic surface's shape.
In the random case, however, atoms corresponding to the same
$\hperp$ can have different local environments and hence 
different parallel-space displacements.  This manifests itself
as split positions in the physical-space cut, or equivalently
the atomic surface becomes an discrete family of surfaces, displaced
from each other by small offsets in parallel space, and 
each having its own distribution $\rho(\hperp).$

Crystallographers would sometimes argue that a
Fourier map reconstructing the correct ensemble-average
density {\it is} ``the structure'' and is the proper
goal when we fit the diffraction. In other words, 
they would say the task is done as soon as the phase problem is solved. 
We disagree strongly with this viewpoint. 
The aim of structural studies is to uncover
the actual structure, which means understanding
what a typical realization is like and not just
the average over realizations.
If the averaged structure contains two nearby, 
half-occupied sites, does the actual structure 
contain exactly one atom which can occupy either
site at random?  Or is it that $50\%$ of the time
{\it both} sites are occupied, and at other times
some other site is occupied?  Spurious but rare
sites have negligible effect on the $R$-factor, 
but major effects on the total energy 
and the electronic states computed from the structure. 
Hence, the goal should be a proper description of the
entire ensemble of configurations. 

\subsection {Outline of the paper}

There are two alternate approaches for solving a random tiling
structure:

{\em 1. ``Standard'' approach}. Break up the task into two stages. 
The first stage is to solve the phase problem 
without a complete model of the real-space structure. 
The (averaged) scattering density $\rho(\r)$ can then be
constructed as an incommensurate cut in a high-dimensional space.
The second stage is to relate this density to an atomic model. 

{\em 2. ``Unified'' approach}. 
Assume a particular kind of random-tiling model, with 
variable parameters (such as the positions and species of
decorating atoms on the tiles, or coefficients in the 
``tile Hamiltonian'' that governs the statistics of patterns
in the random tiling).  Simulate the tiling, calculate
the Bragg intensities, and calculate an $R$ factor to measure
its deviation from the measured intensities. 
Adjust the parameters in a direction that
will reduce $R$ and repeat the simulation; iterate until
a good fit is achieved. 
Following this path, the phase solution is unified with
the structure fitting; it is also natural to unify
the $R$-factor from diffraction and the total energy
in a combined objective function to be minimized by the fit.

Most of this paper is devoted to elaborating the two approaches.
We begin with the ``standard'' path in sec.~\ref{sec:direct-veit}, 
focusing on the new ``minimum-charge'' method, a 
direct method for determining the phases of a general structure.
This method is not peculiar to quasicrystals, nor to random
structures.
However, the minimum-charge viewpoint is useful for random
structures since it provides an
inequality for the diffuse scattering (Sec.~\ref{sec:inequalities})
which allows us to quantify the degree of disorder without 
solving the structure, even partially.
Also, preliminary calculations using 
the minimum-charge approach have
suggested the structure of $i$(AlPdMn)
is more disordered than any of us had believed.

As we have just argued, even a perfect solution of the 
phase problem is very far from giving a useful model of the
structure.
This gap can be bridged by the brutally simple 
``factorization approximation'': assuming that
the intensities are those of ideal structure 
apart from a (Gaussian) ``perp Debye-Waller'' factor.
This is critiqued and 
tested by simulations in 
Sec.~\ref{sec:factorization}.

After Sec.~\ref{sec:factorization}
we turn to the ``unified'' path.
We first review
key notions of tile-decorations
(Sec.~\ref{sec:review}),  since these
are used to specify random-tiling structures,  
practically by the definition of a random tiling.
Then Sec.~\ref{sec:unifit} describes several fitting
procedures, 
beginning with a recipe for discovering the appropriate 
random tiling model by simulations (provided a set of
interatomic pair potentials is known).
Finally, we explain 
in Sec.~\ref{sec:decagonal} the
special problems posed by decagonal structures.

\rem{ Rem2. IS THIS NOW  THE RIGHT PLACE FOR THIS PARAGRAPH?}
Our subject is {\it not} 
the long-wavelength ``phason'' elasticity which produces 
diffuse wings around the Bragg peaks.
These phenomena indeed characterize 
``random tiling'' type behavior, but 
as argued previously in Ref.~\onlinecite{Hen91ART}, 
one should {\it not} necessarily
think this accounts for most of the disorder. 
A priori, one expects that the 
disorder is mostly correlated at the scale of a tile edge or so;
these correlations are controlled by the constraints
of the tiling rules, and produce 
diffuse scattering which is not associated with particular Bragg peaks. 

\section {Solving the phase problem}
\label{sec:direct-veit}

The ``standard'' approach to structure solution
begins by first solving the phase problem. 
This step is somewhat different than for ordinary crystals, 
but it seems not to matter much whether 
the quasicrystal is modeled as ideal or as a random tiling.

We first mention a couple of well-known techniques.
An old technique~\cite{Cor91} takes advantage of the 
smooth dependence of the structure factor $F_\GGpar$ when plotted
as $F(\GGperp)$, where 
$\GGperp$ is the perp-space partner of $\GGpar$.
Where $|F(\GGperp)|$
passes through a zero, one infers that $F(\GGperp)$ must
change sign.  
Another approach, described by Qiu and Jari\'c, 
was to discover phases using known rational approximant
crystals (by expressing them as rational cuts of a 
higher-dimensional structure).~\cite{Qiu90}. 

In the rest of this section, we first outline a new
approach, the ``minimum-charge'' method (Sec.~\ref{sec:minQ}).
Quite independent of this, we derive an  inequality
which is useful as a diagnostic of disorder (Sec.~\ref{sec:inequalities}). 
Finally, we apply both of these ideas to a data set for $i$(AlPdMn), in
Sec.~\ref{sec:boudard}.

\subsection {``Minimum-charge'' approach}
\label{sec:minQ}

A promising new ``direct method'' (not specific to quasicrystals)
to solve the X-ray diffraction phase problem
was suggested by one of us\cite{Els99}.
The key notion of this is to
minimize the average electron charge density. 
Any periodic or quasiperiodic density has the form
  \begin{equation}
     \rho(\r) = 
      F_0 + \sum_{\GGpar\neq 0} |F_\GGpar| \cos(\GGpar\cdot \r -\phi_\GGpar),
  \label{eq:veit1}
  \end{equation}
where $F_0$ is the average density; $|F_\GGpar|\propto \sqrt{I_\GGpar}$ 
and $\phi_\GGpar$ are the magnitudes 
and phases of structure factors.
The former are obtainable from the measured intensities $I_\GGpar$
and the latter are to be determined.
The ``minimum charge" method\cite{Els99} considers the reduced density
  \begin{equation}
    \tilde{\rho}(\r) \equiv \rho(\r)-F_0,
  \end{equation}
and seeks phases $\phi_\GGpar$ which maximize the minimum value of
$\tilde{\rho}(\r)$ 
      \begin{equation}
      \tilde{\rho}_{\rm min}=\min_{\r}\tilde{\rho}(\r).
   \end{equation}
The minimum value of $F_0$ which makes $\rho(\r)$ everywhere
nonnegative is then $F_0=-\tilde{\rho}_{\rm min}$. It is
straightforward to argue that at an optimal set of $N$ phases (where
$F_0$ has achieved a local minimum) the minimum $\tilde{\rho}_{\rm
min}$ is attained at $N+1$ exactly degenerate local minima of
$\tilde{\rho}(\r)$. Consequently, for optimal phases, the minimum
``background" value of the density is found in large regions of the
unit cell. If some phase assignment allows for the volume occupied by
this background density to be especially large, as we expect when
dealing with a true atomic density, the corresponding value of $F_0$
will be especially low. This property distinguishes the global
minimum of $F_0$ and provides for an unambiguous phase solution.

By sampling $\tilde{\rho}(\r)$ on a ``Fourier grid"\cite{Elser00} of a
comparable size, the charge minimization problem, say for a 
centrosymmetric space group,
is cast into a mixed integer
programming optimization\cite{optGenRef} for the unknown (integer)
signs and (real valued)
 $F_0$. Because of linearity (of objective
function and constraints) there are efficient search strategies for
this general class of problem. For the $i$(AlPdMn) 
x-ray phases (see Subsec.~\ref{sec:boudard})
the ``branch and bound'' technique was used, 
where the relaxation of the constraint
$|s_\GGpar|=1$ (on each of the signs) to the weaker statement $|s_\GGpar|\leq 1$
(on a subset of the signs) provides bounds on the objective function
($F_0$) that frequently are strong enough to eliminate large branches
of the search tree.

There is a similarity between the minimum charge ({\it minQ}) and the
maximum entropy ({\it maxS}) methods. Both succeed in eliminating or
minimizing wiggles in the electron density in regions where there
should be no charge. The similarity is only superficial, however.
On the one hand, {\it maxS} is fundamentally a strategy for refining phases
 which are already approximately determined, but  corrupted by noise;
 on the other hand, {\it minQ} makes perfect sense in the context of
perfect data. Also, {\it minQ} was developed specifically for {\it ab
initio} phase determination, whereas the application of {\it maxS} in
this mode has no theoretical basis.

\subsection {Inequalities for X-ray structure factors}
\label{sec:inequalities}

The aim of this subsection is 
to place a lower bound on the amount of disorder, taking  advantage
of the fact that the density of scattering electrons 
is nonnegative.
This is completely independent of 
the preceding section and, in particular, 
all the conclusions of Subsec.~\ref{sec:inequalities}
can be drawn without ever determining or even considering 
a single phase factor. 

The positivity of $\rho(\r)$ implies the inequality
   \begin{equation}
      |F_\GGpar|=|\left\langle\rho(\r) e^{-i \GGpar\cdot \r}\right\rangle_{r}|
       \leq
      \left\langle|\rho(\r)|\right\rangle_{r}=F_0,
   \label{eq:veit7}
   \end{equation}
where $\langle \ldots \rangle_r$ means a {\it spatial} average.
It should be emphasized that $\rho(\r)$ already contains 
an {\it ensemble} average:
that is just what the Bragg scattering represents, as we
noted in Sec.~\ref{sec:perpspace}.
Eq. (\ref{eq:veit7})
is the most trivial of the Harker-Kasper inequalities~\cite{Harker}. 
Note, though, that in crystallography, these (and later inequalities) 
were always applied to the solution of the
phase problem assuming an ideal structure. 
We apply (\ref{eq:veit7}) instead as a rigorous diagnostic 
of a {\it disordered} structure, using only Bragg (not diffuse) scattering.

It follows very obviously from (\ref{eq:veit7}) that
      ${|F_\GGpar|}/{\sqrt{\sum |F_\GGpar|^2}}$
      $\leq {|F_0 |}/{\sqrt{\sum |F_\GGpar|^2}}$
which is conveniently written
   \begin{equation}
      \frac{|F_\GGpar|}{\sqrt{\sum _{\GGpar \neq 0} |F_\GGpar|^2}}
           \leq  \alpha
   \label{eq:veit8}
   \end{equation}
where 
   \begin{equation}
         \alpha \equiv      \frac{F_0}{\sqrt{\sum_{\GGpar\neq 0} |F_\GGpar|^2}}.
   \label{eq:alpha}
   \end{equation}
Note that $\alpha$ just depends on the first two moments of $\rho$, since
   \begin{eqnarray}
   \label{eq:veit6a}
      \left\langle\rho\right\rangle_r &=&F_0, \\
   \label{eq:veit6b}
      \left\langle\rho^2\right\rangle_r &=&F_0^2+\sum_{\GGpar\neq 0}|F_\GGpar|^2.
   \end{eqnarray}

The value of (\ref{eq:veit8}) is that both
sides can be related to experimentally measurable data, as is
evident for the Bragg intensities in the 
left hand side. (The sum omits $F_0$, 
which is not measurable as a Bragg intensity.)
The right hand side 
turns out to depend only on atomic parameters such as charges, number density, 
concentrations, etc., which are known from the composition, 
{\it provided we assume the 
diffracting material consists of a known density of definite atomic 
species} (well separated so overlaps of atomic charge distributions 
are negligible). 
Specifically, suppose there is no disorder other than that due to 
thermal vibrations of the
atoms. To a good approximation one may then represent the electron
density as a sum of non-overlapping distributions centered around
each atom, $\rhat_i(\r)$ where $i$ is the atomic species. Without knowing the
atomic positions, it is possible to evaluate the averages in 
(\ref{eq:veit6a}) and (\ref{eq:veit6b}):
   \begin{eqnarray}
   \label{eq:veit4a}
      \left\langle\rho\right\rangle_r &=&n\sum_{i}x_i Q_i\\
   \label{eq:veit4b}
      \left\langle\rho^2\right\rangle_r &=&n\sum_{i}x_i P_i,
   \end{eqnarray}
where $n$ is the number density of atoms, $x_i$ the atomic composition, and
   \begin{eqnarray}
      Q_i &=& \int \rhat_i(\r) d^3 \r  \\
      P_i &=& \int \rhat_i(\r) ^2 d^3 \r.
   \label{eq:veitP}
   \end{eqnarray}


Inserting (\ref{eq:veit4a}) and (\ref{eq:veit4b}) into the
earlier equations, 
(\ref{eq:veit6a}) and
(\ref{eq:veit6b}), 
we conclude $\alpha = \alpha_0$, where
   \begin{equation}
      \alpha_0 \equiv \frac{1}{\sqrt{\frac{\sum x_i P_i}{n(\sum x_i Q_i)^2}-1}}.
      \label{eq:veit8a}
   \end{equation}
is expressed purely in terms of the
atomic parameters $n$, $x_i$, $Q_i$ and $P_i$,

Our inequality (\ref{eq:veit8}) is rigorous. If it is violated
by the value of $\alpha_0$ computed from the data, 
it means we were mistaken in our assumption
that the electron density $\rho(\r)$
has 100\% occupation of every site and no
chemical disorder. 
Let us repeat the derivation above, assuming a distribution of
occupations (in the case of a single species for simplicity).
In place of (\ref{eq:veit8a}) we find
   \begin{equation}
      \alpha \simeq \sqrt{\frac{n Q^2}{P \langle y\rangle_{z}}},
   \label{eq:alphadist}
   \end{equation}
in the case where $\alpha \ll 1$. Here $z(y)$ is the distribution of
the fraction of charge from sites with fractional occupation $y$. 
Disorder decreases $\langle y \rangle$ and thereby, 
according to (\ref{eq:alphadist}),  increases $\alpha$
beyond the value in (\ref{eq:veit8a}). 
If we allow for enough occupational disorder, 
inequality (\ref{eq:veit8}) will be satisfied. 

\subsection {Example: the $i$(AlPdMn) density map}
\label{sec:boudard}

Recently\cite{AvanesovBrownElser} the ``minimum-charge'' method
was applied to determine the phases for $i$(AlPdMn),
which has the centrosymmetric space group
$F\bar{5}\bar{3}\frac{2}{m}$, 
from the x-ray data of Boudard et al.\cite{Bou92}.
This data set has 503 symmetry
inequivalent reflections, 360 having $\sigma_\GGpar/I_\GGpar <0.5$;
we used all the reflections. 

\begin{figure}
\begin{minipage}{15cm}
\hbox { \epsfxsize=7.0cm \hskip  0in  \epsfbox{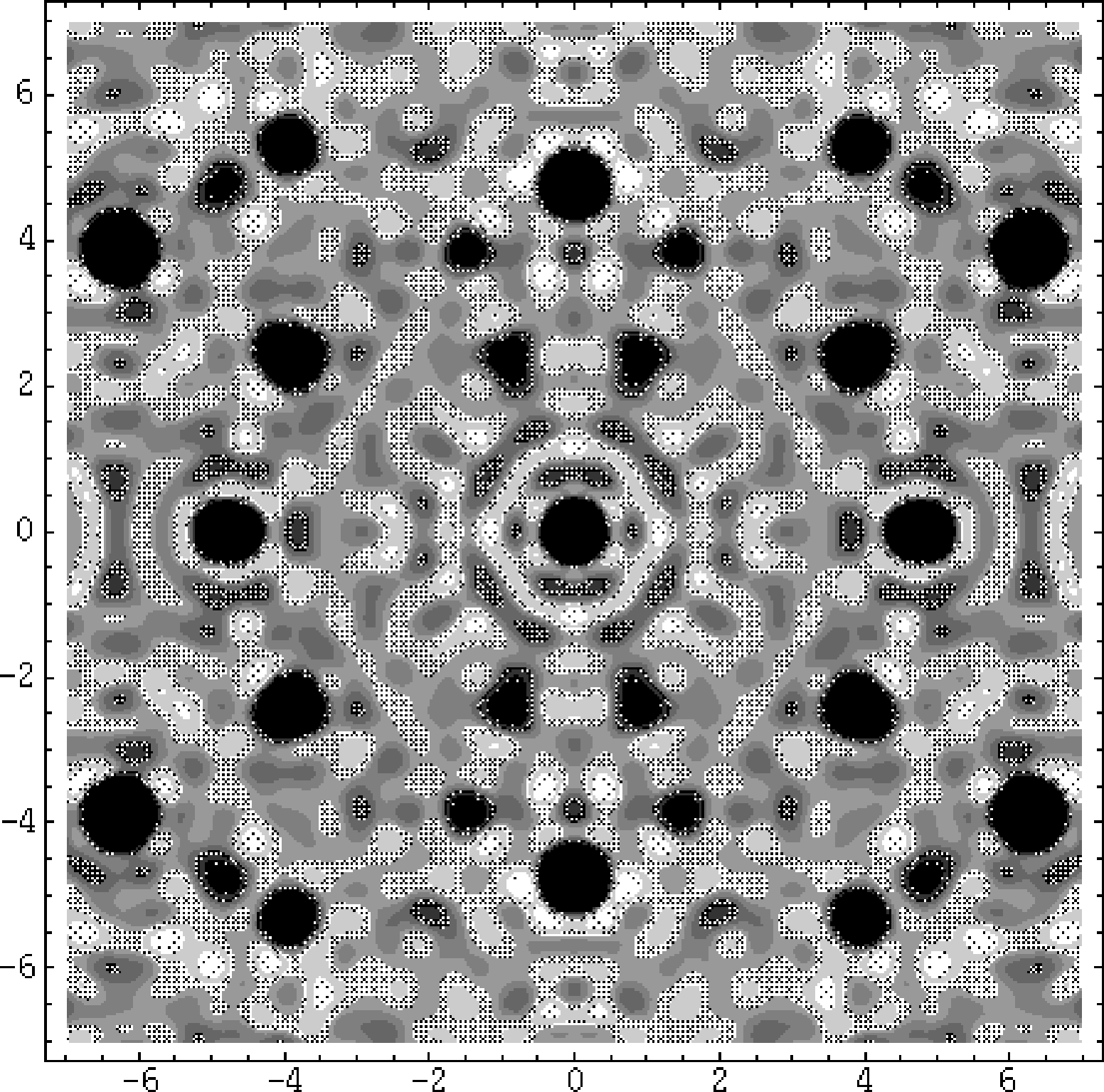}
          \hskip 0.25in \epsfxsize=9.5cm \epsfbox{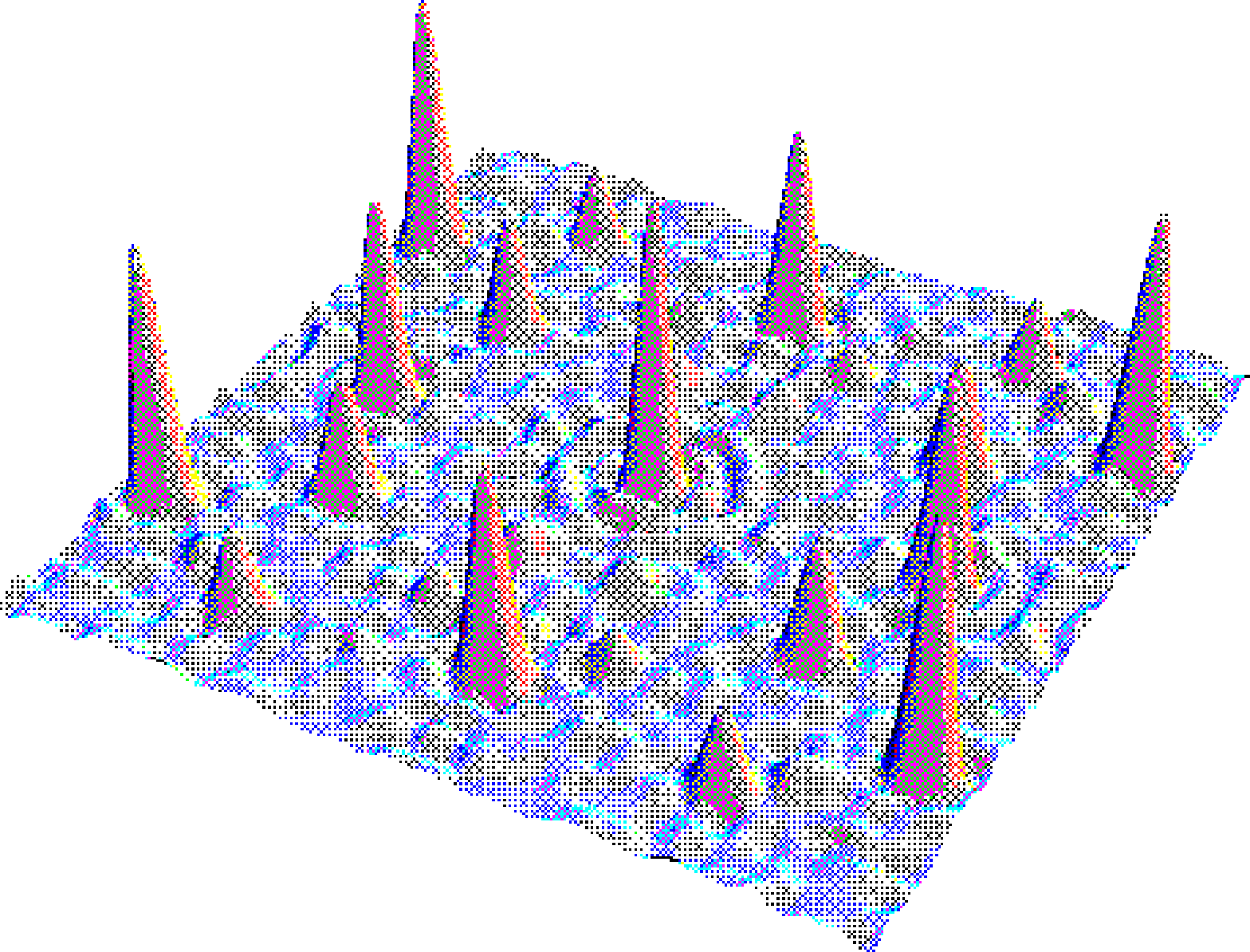}}
\caption
{Average electron density $\rho(\r)$ 
of $i$(AlPdMn) in physical space, reconstructed from
503 
experimental x-ray structure factors of \protect\onlinecite{Bou92}
using signs obtained by the minimum
charge algorithm of Sec.~\protect\ref{sec:direct-veit}.
The plane of the plot is an icosahedral mirror plane
($13\AA\times 13\AA$) centered on a pseudo-Mackay cluster 
($n_0$ atomic surface).
Charge density is represented by darkness 
in (a) and by height in (b). 
There are several instances of atoms (or partial atoms)
at unphysically short separation.}
\label{fig:mackay}
\label{fig:rho}
\end{minipage}
\end{figure}

\begin{figure}
\begin{minipage}{15cm}
    \hbox{\epsfxsize=7.0cm \hskip 0in \epsfbox{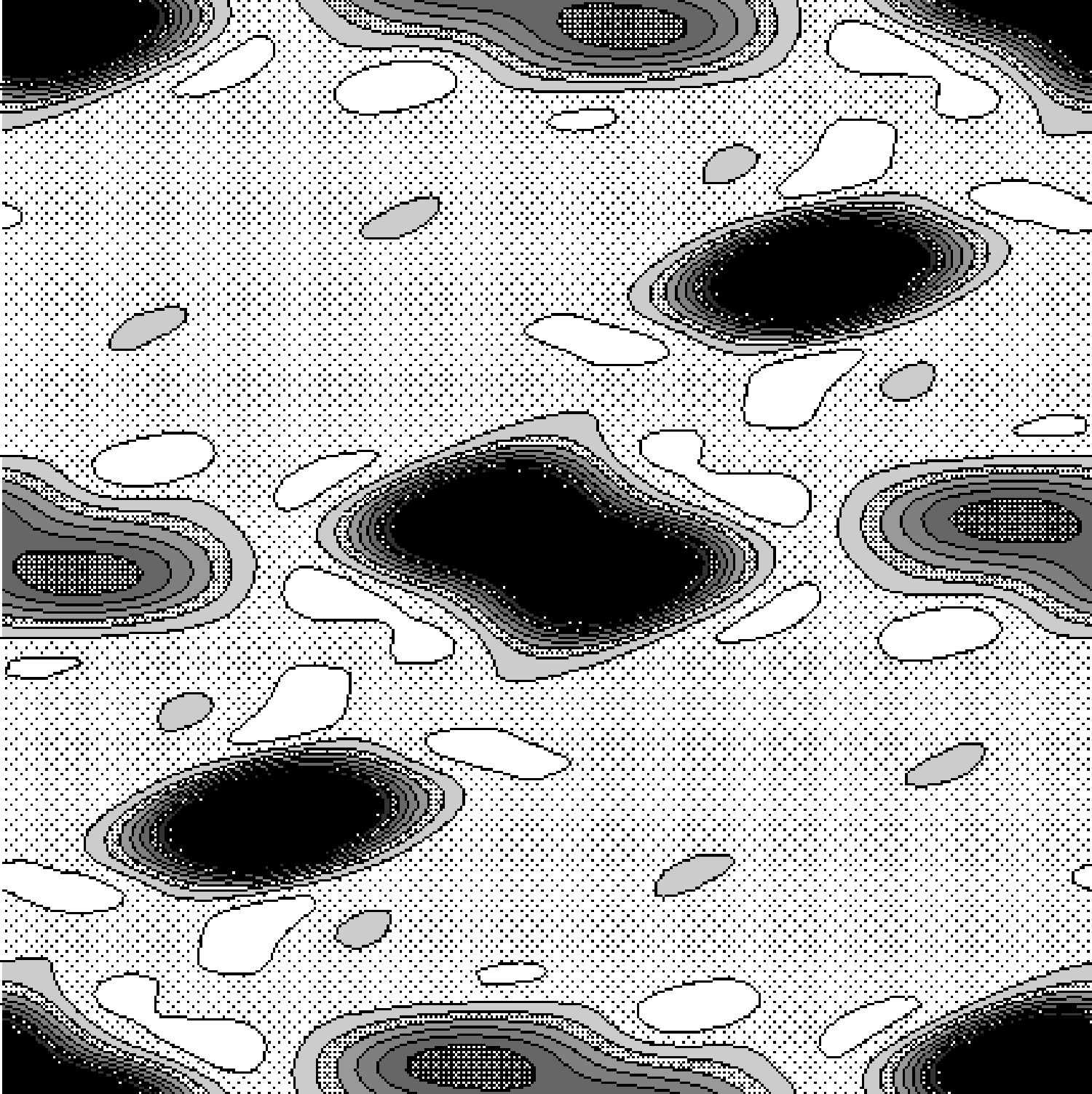}
    \hskip 0.25in \epsfxsize=9.5cm \epsfbox{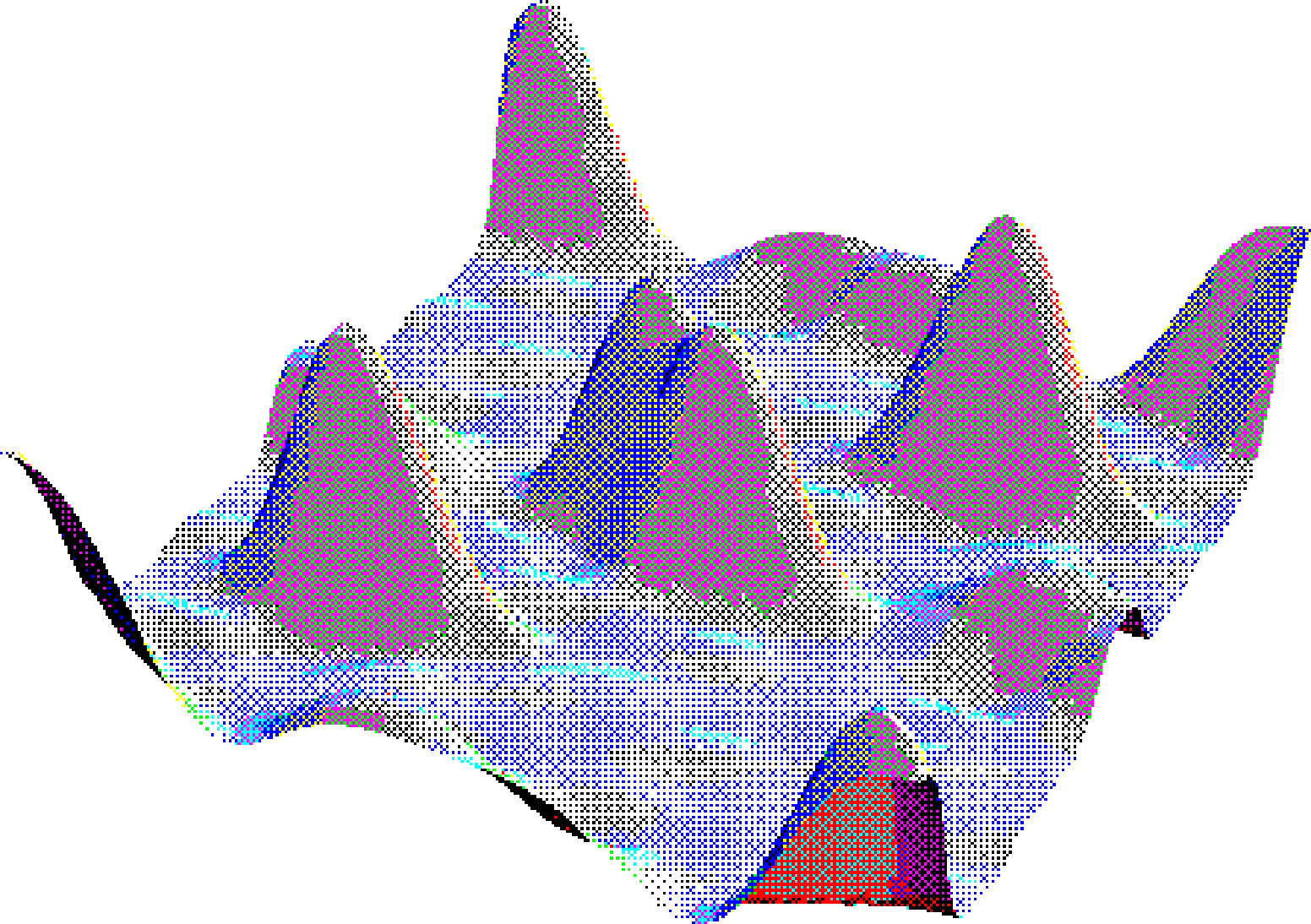}}
\caption
{The same electron density plotted in
Figure \ref{fig:mackay} but for the 5-fold periodic plane.
The figure box is a unit cells of this plane, so
the physical and perp space directions run obliquely
\VEIT{Can you clarify this?}
The $n_0$
surface is centered at $(0,0)$ and $(1/2,1/2)$, $n_1$ at $(1/2,0)$ and
$(0,1/2)$, and ${bc}_1$ at $(1/4,1/4)$ and $(3/4,3/4)$. A $\tau^3$
inflation has been applied to both plots to make the rounding of features
in parallel and perp-space comparable in magnitude (i.e. so the respective
Debye-Waller factors are similar); 
this is seen most clearly in the contour plot (a). The height
plot (b) shows that the density profile of the  $n_0$ ``surface'' even in
the perp-direction, 
is rounded to an extent comparable to its entire width.}
\label{fig:periodic}
\end{minipage}
\end{figure}

Figures \ref{fig:mackay} and
\ref{fig:periodic} both
show the reconstructed electron density using signs
obtained with the minimum charge method. 
Slices through physical space (Fig.~\ref{fig:mackay}),
and along rational planes (Fig.~\ref{fig:periodic}), 
both show disorder, as we argue in the next two paragraphs.
Disorder of this sort is expected in a random tiling, 
but our observations here do not rule out other origins.

Figure \ref{fig:mackay}(a) shows the electron density in a 
2-fold (mirror) plane of physical space, centered on the ``$n_0$'' atomic
surface. While a subset of the atoms corresponds
exactly to the Mackay cluster second shell  (12+30 atoms),
we see that there are
also several pairs of well defined atoms
having unphysically short separations. Similar evidence of split
positions has been noted before by others, but was never ruled
definitive because of the possibility of being a truncation
artifact\cite{Bois-trunc}. Because of the flatness of the background in
these reconstructed densities we feel that truncation cannot be
invoked to explain these examples of split positions. A preliminary
survey of the reconstructed density suggests that split positions are
actually quite common.

In the periodic 5-fold plane (Fig.~\ref{fig:periodic}) we see
three atomic surfaces centered at Wyckoff positions with
icosahedral symmetry. The positions and net charge in these
surfaces are consistent with the ``spherical model'' of
Boudard {\it et al}\cite{Bou92}. 
We observe, however, that the density profiles of all three
surfaces are very rounded, rather than step-function-like.~\footnote{
We caution that 
some important details are lost when the weak reflections are omitted;
in other words, truncation error may introduce a spurious 
apparent randomness.
For example, with only 300 reflections the 
profile of the atomic surface at the center of 
Fig.~\ref{fig:periodic}(b) looks quite Gaussian.
(Its correct appearance is two-peaked 
because its center in perp space is
occupied by Mn, while the rest of the atomic surface
is Pd which possesse twice as many electrons as Mn.)}
The width of the appropriate broadening function in the
factorization approximation
(see eq.~(\ref{eq:convolve}), below) 
would be comparable in magnitude to the diameter of
the largest (ideal) atomic surface in the spherical model.
This explains, for example, why the sizes of the largest
(``$n_0$'') and smallest (``$bc_1$'') surfaces of
Ref.~\onlinecite{Bou92} are comparable in
our image, whereas in the spherical model their diameters
are in the ratio 2.3:1. 

Even before determining the phases, 
the Boudard data\cite{Bou92} can be manipulated to show (using the
inequalities in Sec.~\ref{sec:inequalities}) that the $i$(AlPdMn) 
phase possesses considerable partial/mixed occupational disorder. 
To evaluate $\alpha$ from eq.~(\ref{eq:veit8a}),
we first approximated the atomic form factor 
for each species $i$ as a Gaussian,
   \begin{equation}
      \rhat_i (\qq)=
Q_i e^{-\frac{1}{2}B_i |\qq|^2},
   \end{equation}
with the parameters $B_i$ chosen to reproduce the integrals $P_i$
in (\ref{eq:veitP}) 
obtained using Hartree-Fock wavefunctions. This gave $B_{\rm
Al}=0.012$, $B_{\rm Pd}=0.0042$, and $B_{\rm Mn}=0.007$ (in units of
${\AA}^{2}$). Since
$B_{\rm Pd}$ is close to the measured Debye-Waller factor for
Pd\cite{Bois94a}, $B_{\rm DW}=0.0044$, the thermally averaged
distribution for Pd is better represented by a Gaussian with
$B=0.0086$. Since Pd atoms make the main contribution to $\alpha$, we
used the same Debye-Waller factor to correct the Al and Mn
distributions. Using the measured composition and
density\cite{Bou92}, we finally arrive at $\alpha_0= 0.0439$.

\rem {Rem4. WAS REWORDED AFTER VEIT LOOKED.}
The sum $\sum|F_\GGpar|^2$ in (\ref{eq:veit8})
is in principle determined by a measurement
of all the Bragg intensities. 
Barring surprises in the unexplored regions of reciprocal space, 
we believe the intensities that have been measured exhaust this sum: 
the weaker measured peaks of our data set already make a negligible
contribution to it, and peaks with larger $\GGpar$ or $\GGperp$
are cut off by Debye-Waller factors.
Our conclusion about the disorder follows from the fact
If we assumed no disorder, then $\alpha=\alpha_0$ and
the three most intense reflections (18/29,52/84, and 20/32) would violate
inequality (\ref{eq:veit8})
by as much as 40\%. 
Our estimate of disorder is
certainly conservative since there is no reason to expect that
inequality (\ref{eq:veit7})
would be close to an equality.
(Sec.~\ref{sec:perpdist} also indicates a large
disorder in this material.)

\section {Factorization approximation}
\label{sec:factorization}


The determination of the phases, and hence of the averaged
scattering density, is not the end of the structure determination
in the random-tiling case.  A process of simulation and fitting is
still called for. (See Sec.~\ref{sec:unifit}, below.)
The factorization approximation -- equivalently the 
``Phason  Debye-Waller factor'' -- is a shortcut of dubious validity, 
which is nevertheless attractive since it offers the hope of relating
the data to an ideal model with much less effort.

In fitting diffraction, it was natural to generalize the 
Debye-Waller factor to perp space and to assume
    \begin{equation}
         F^\rand_\GGpar = 
         F^\id_\GGpar e^{-W(\GGpar)}
    \label{eq:DWreduced}
    \end{equation}
for the Fourier amplitudes of the random and ideal tilings
(labeled by ``$\rand$'' and ``$\id$'' henceforth).
Here the ``perp Debye-Waller factor'' is given by
    \begin{equation}
    W(\GGpar) = \sigma^2 |\GGperp|^2 /2
    \label{eq:DW}
    \end{equation}
where $\GGperp$ is the perp-space wavevector corresponding
to $\GGpar$.
We will call this the ``factorization approximation''.
In direct (perp) space, in view of (\ref{eq:FQFT}), 
eq.~(\ref{eq:DWreduced}) is equivalent, for an icosahedral
quasicrystal, to
    \begin{equation}
         \rho^\rand(\hperp) = 
         \int d^{3} \uperp
          w(\uperp) \rho^\id(\hperp-\uperp)
    \label{eq:convolve}
    \end{equation}
where $w(\uperp)$ is a (normalized) smearing
function with a Gaussian profile:
    \begin{equation}
          w(\uperp) = 
          (2 \pi \sigma^2)^{-3/2} \exp[-|\uperp|^2/2\sigma^2]
    \label{eq:smearing}
    \end{equation}
Notice that it follows from (\ref{eq:convolve}) that
the random tiling vertices have a perp variance increased
over the ideal tiling by 
    \begin{equation}
       \langle |\hperp|^2 \rangle_\rand - 
       \langle |\hperp|^2 \rangle_\id = 3 \sigma^2 .
    \label{eq:perpvardiff}
    \end{equation}
For a general quasicrystal, ``$3$'' must be replaced by
``$d_\perp$'' in 
eqs.~(\ref{eq:convolve}) -- (\ref{eq:perpvardiff}).
(We wrote them assuming perp space is three-dimensional as
in the icosahedral case.)

It must be borne in mind, however, that there is {\it no} 
exact basis for the factorization approximation.
One may be misled because elastic theory tells us all {\it long 
wavelength} perp fluctuations are Gaussian.~\footnote{
This does imply a formula like (\ref{eq:convolve}) relating
$\rho^\rand(\hperp)$
to the coarse-grained distribution $\rho^{\rm short}(\hperp)$, 
obtained after averaging the ``center-of-mass'' in perp space
locally, from a region several tiles  wide,
rather than over the entire system.}
But this relation is not useful, 
since we expect much of the random-tiling disorder 
to be short-range.

Another way one may be misled is that, when the higher-dimensional
cut construction is overemphasized, a natural cartoon of the
random tiling is to take the {\it same} higher-dimensional crystal
as in the ideal case, 
but to permit the cut-surface to undulate --
to deviate from being a flat plane.
If the cut-surface has Gaussian fluctuations that are completely
uncorrelated with the atomic surfaces, then the factorization
approximation would become exact.  
This ``undulating cut''
notion was critiqued in Sec.~4 of Ref.~\onlinecite{Hen91ART}.\footnote{
In places where a straight cut would always cross exactly one out of 
two atomic surfaces, 
an undulating cut may cross both or neither  of them, 
which produces an overlap or gap between parts of tiles.
Additionally, two slightly different undulating cuts often
produce the same real structure, so one must be careful
in assigning a statistical weighting to the ensemble of undulating
cut surfaces. 
}

The factorization assumption is fundamentally ill-defined since
the same random tiling (e.g., of Penrose rhombi) may be
obtained as the randomization of {\it various} different quasiperiodic
tilings (e.g. the generalized Penrose tilings in two dimensions) --
yet eq.~(\ref{eq:DWreduced}) demands that we identify 
a {\it particular} ideal tiling as the one which has been randomized.
The physical meaning of (\ref{eq:convolve}) is dubious, too:
it says that, when the tiling is randomized, the
vertices with a prescribed local environment, which in the ideal
tiling have a specific $\hperp$ value, now get shifted by a
random offset $\uperp$ which has a probability distribution
(\ref{eq:smearing}).  But there is no reason the random variable
$\uperp$ should have
the same variance (\ref{eq:perpvardiff}), independent of
what kind of local environment is in question or 
of how large is its $\hperp$ location.

We claim the factorization approximation works 
trivially for sufficiently small $\GGperp$.
After all, an integral of form (\ref{eq:FQFT}) can always
be written as $\exp (- C_0 + \frac{1}{2} C_2 |\GGperp|^2 +  \ldots)$, 
where $C_{2k}$ is the {\it cumulant} of order $2k$ of the density
$\rho(\hperp)$.
Then since the $2k=4$ term in the exponent is of order $(\GGperp)^4$, 
it follows that $F(\GGperp)$ has an  roughly Gaussian shape at small $\GGperp$, 
even for the ideal tiling (see Ref.~\onlinecite{Els85b}).  
The real test of the factorization approximation is
whether it works for larger $\GGperp$, 
with the fitted $\sigma^2$ in (\ref{eq:DW})
being independent of $\GGperp$. 
Surprisingly, this seems to be true for our simulations,
as the rest of this section shows.

\subsection {Simulation tests}
\label{sec:simdetails}

We carried out simulations of the
completely random tiling of 
rhombohedra~\cite{Tang90,Shaw91,Mih99}, 
as a toy model to test (for the first time)
the relation between diffraction amplitudes in the
ideal and random tilings: in particular, 
to test the validity of Eq.~(\ref{eq:DWreduced}). 
These measurements are preliminary; we hope they
can be repeated in the future, more systematically and
perhaps on models closer to the real  quasicrystals.
The rest of this section is devoted to these simulations.

We used the cell of the ``8/5'' cubic approximant with
periodic boundary conditions, which contains 
10 336 vertices (an equal number of rhombohedra). 
For equilibration we allowed 500 Monte Carlo steps (MCS) / vertex
(these are flip attempts of which only $17\%$ cause flips), and
then our ensemble for diffraction consisted of 
1000 sample taken at intervals of 250 MCS/vertex.
(This is adequate to decorrelate this maximally random model,~\footnote{
The relaxation time of the 
slowest Fourier mode in the 8/5 cell was found to be 
only 280 MCS/site~\cite{Shaw91}. Our Fig.~\ref{fig:braggdiffuse}
agrees well with Fig.~1 of Ref.~\onlinecite{Tang90}; 
furthermore, a different run with the
same protocol yielded elastic constants $(K_1,K_2)=(0.84,0.50)$
within 2\% of the results for the same size in Ref.~\onlinecite{Shaw91}, 
Fig.~2.}
 
but quite inadequate when there is a Hamiltonian
\cite{Mih99}.)

A novel detail of our simulation is
that periodic boundary conditions are also assumed, rather
arbitrarily, in the 
{\it perp} direction. Thus, only a finite set of
$(178)^3$ $\hperp$ values is possible.
In practice all vertices fall within a smaller domain
of these in perp-space -- $130^3$ points 
were used here -- and configurations 
are efficiently represented as a 
lattice gas on this grid.

\subsubsection{Results on perp-space distribution}
\label{sec:perpdist}

It is possible to visualize the smearing of 
eq.~(\ref{eq:convolve})
directly, rather than through the perp Debye-Waller factor.
On the theory side, 
it is straightforward to construct a histogram of $\hperp$
values for the vertices  of the
random rhombohedral tiling (see Sec.~\ref{sec:simdetails}, below).
A random-tiling decoration model~\cite{Els97} has been proposed for
$i$(AlCuFe) and $i$(AlPdMn). This is based on the rhombohedral
random tiling, which has simple-icosahedral space group. However, the
atomic decoration treats even and odd vertices inequivalently, 
so the atomic structure is face-centered icosahedral, 
like the real quasicrystals, 
and is similar to prior models based on diffraction~\cite{Cor91,Bou92}
In particular, the $bc_1$
atomic surface is (in this model) made up precisely of atoms that decorate 
the {\it even} rhombohedron nodes and should have the same
perp-space distribution $\rho(\hperp)$ 
as {\it all} nodes do in the simulation (since even and odd nodes have
the same $\rho(\hperp)$ in this model). 
In short, the experimental perp-space distribution of the $bc_1$ surface
is just a section through the density map 
(Sec.~\ref{sec:direct-veit}).

\begin{figure}
\begin{minipage}{8.66cm}
\epsfxsize=8.66cm \epsfbox{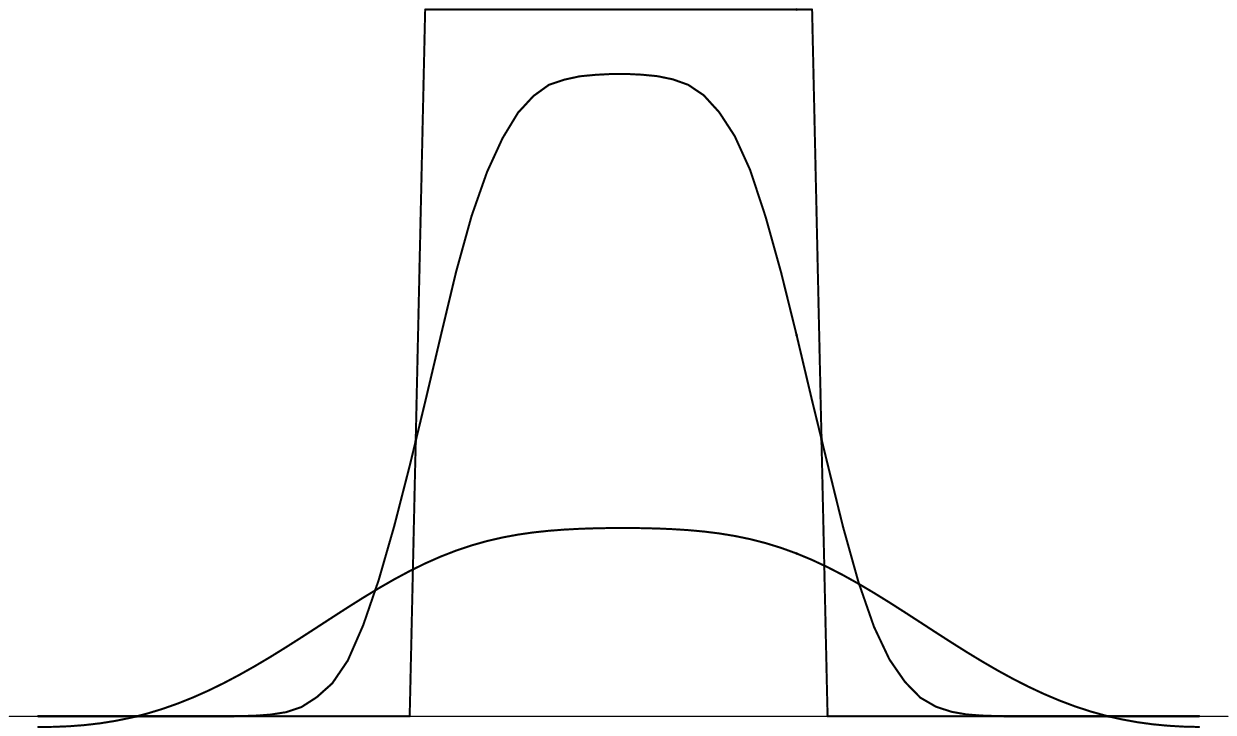}
\caption
{Cross section $\rho(\hperp)$ 
of atomic surface, for ideal and random rhombohedron tilings.
The horizontal axis is a one-dimensional section 
along the twofold direction;
\rem{Veit was supposed to redo this with 503}
the vertical scale is normalized so the 
integral of the density 
$\int d^3 \hperp \rho(\hperp)$,
approximated as spherically symmetric, is the same. 
\rem {Thus the one-dimensional integral of the
curves in this figure are NOT the same.}
The rectangular and
the flat-topped curves are the ideal and random 
rhombohedron tilings, respectively.
The very broad Gaussian is a scan
through the $bc_1$ atomic surface reconstructed 
from Boudard's data~\protect\cite{Bou92} 
using the  minimum-charge method of Sec.~\ref{sec:direct-veit}.
(The horizontal scale is converted to units of 
rhombohedron edges assuming the model
of Ref.~\protect\onlinecite{Els97}.)
This last curve differs from a section of Fig.~\ref{fig:periodic}
in that only 300 reflections are used here.}
\label{fig:densityProfile}
\end{minipage}
\end{figure}

\begin{figure}
\begin{minipage}{8.66cm}
\epsfxsize=8.66cm \epsfbox{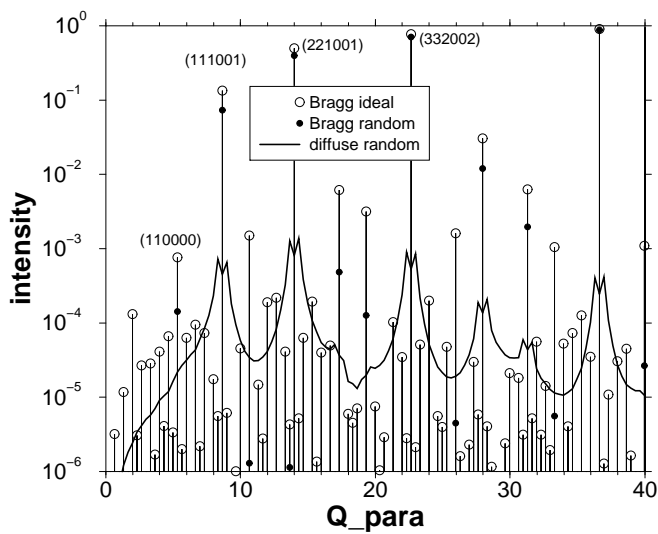}
\caption
{Separate Bragg and diffuse diffraction along
the 2-fold axis  of reciprocal space for the
rhombohedron random tiling (8/5 approximant).
Vertical bars are a guide for the eye and indicate the location
of the Bragg peaks. All reciprocal lattice vectors
are included with intensity over $10^{-6}$.
Some peaks are labelled using Elser's 
convention~\protect\cite{Els85b}. 
The same figure was presented by Tang in 
Ref.~\protect\onlinecite{Tang90}, 
but without separating the Bragg and diffuse parts.}
\label{fig:braggdiffuse}
\end{minipage}
\end{figure}

Surprisingly, assuming the phase reconstruction is correct, 
$i$(AlPdMn) -- the most perfectly ordered quasicrystal to date --
appears ``more random than random'' as quantified by the apparent
value of $\sigma^2$ in Fig.~\ref{fig:densityProfile}.
This conclusion is also supported by the
quite small numerical  values of elastic constants in $i$(AlPdMn),
as measured from diffuse scattering~\cite{LeT99}, in 
comparison to models~\cite{Mih99}.
We know only two possible explanations of this disorder, neither
of which is very convincing:

\begin{itemize}
\item[]
{\it Explanation (i):} 
The true tiles are deflated 
compared to the assumed ones by some factor like $\tau^2$
(such rescalings are discussed in Sec.~9.2.1 of Ref.~\onlinecite{Hen91ART}).
But for $i$(AlPdMn) this seems impossible, as the true tile edge
would be no larger than an interatomic spacing.

\item[]
{\it Explanation (ii):} 
The ratio of elastic constants is close to $K_2/K_1=-0.75$
at which a ``phason instability'' occurs, with a divergence of
perp fluctuations~\cite{diffuse91}.
Within the entropic-stabilization scenario, such an instability is
indeed expected when the temperature is lowered~\cite{entropic} -- but
only near a critical temperature, not over a wide range.
\end{itemize}

Further understanding awaits a better understanding of the
currently unsettled experimental situation. In particular,
Ref.~\cite{Cap99} found that the diffuse scattering
differed by a factor of 20 between two slightly different samples
of $i$(AlPdMn); it is unclear which of these
is like the sample of Ref.~\onlinecite{Bou92}. 

\subsubsection{Computing Bragg and diffuse scattering}

The rest of our studies in this section
depend on the diffraction.
To compute this, each vertex was taken to be a point scatterer
with form factor unity. 
In the size 8/5 approximant, computing the
Fourier transform at {\it every} wavevector
in the interesting range
would take weeks or years of CPU time, so we used lists of 
selected wavevectors (see below). 

When one uses periodic boundary conditions,
of course the real reciprocal-space vectors fall 
on a discrete grid. 
Due to our special adoption (see above) of {\it perp}
periodic boundary conditions, the perp reciprocal-space
vectors also lie on a discrete grid. Indeed 
{\it each} of our real reciprocal-space vectors $\qq$ is the
projection of infinitely many 6D reciprocal-lattice vectors, 
and can thus be considered as a Bragg vector (with the appropriate
distortion since this is an approximant).
We select the smallest of these to make the correspondence unique.
In practice most of the $\qq$ vectors 
correspond to such a large $\GGperp$ that the Bragg amplitude is zero to
computer precision.

We defined the ``Bragg'' and ``diffuse'' parts of the diffraction
according to their behavior in the infinite-size limit, so we need
a new operational definition or procedure to separate
the Bragg from the diffuse intensity in a fixed, finite lattice. 
This is easy to do, in view of the discussion in 
Sec.~\ref{sec:perpspace}:
the amplitude's ensemble average is taken to be the
Bragg  amplitude, while its variance is the diffuse intensity. 
However, even if each sample were very similar to the
previous one, it is offset by a random
fraction of the simulation cell, independent of the
previous offset.
Hence, before taking the Fourier sum over each sample, 
we located the ``center-of-mass'' of its perp-space
coordinates, and then uniformly shifted the vertex coodinates
(by integer multiples of the underlying grid of the lattice-gas)
so as to make this center-of-mass nearly zero. 

In this fashion we could measure Bragg intensity even on peaks
where the diffuse intensity was perhaps $\sim 10$ times as big.~\footnote{
If the system size were changed, then of course the Bragg
{\it amplitude} and the diffuse {\it intensity} both scale linearly
with  the system's volume.}
The results (along a twofold axis) are shown in
Fig.~\ref{fig:braggdiffuse}.
Bragg peaks of the ideal tiling are identifiable
down to amplitude $10^{-4}$ relative to the maximum (far smaller
than experimentally measurable).
Bragg amplitudes of order $10^{-2}$ in the ideal tiling became
of order $10^{-4}$  in the random tiling, which is about where
they become lost in the diffuse noise. 
In the best experimental data sets, the smallest measured Bragg amplitude
is down about $10^{-2}$ from the largest one.

\subsection {Results: perp-space Debye-Waller factor}

\begin{figure}
\begin{minipage}{13cm}
\hbox {
   \epsfxsize=8.4 cm \epsfbox{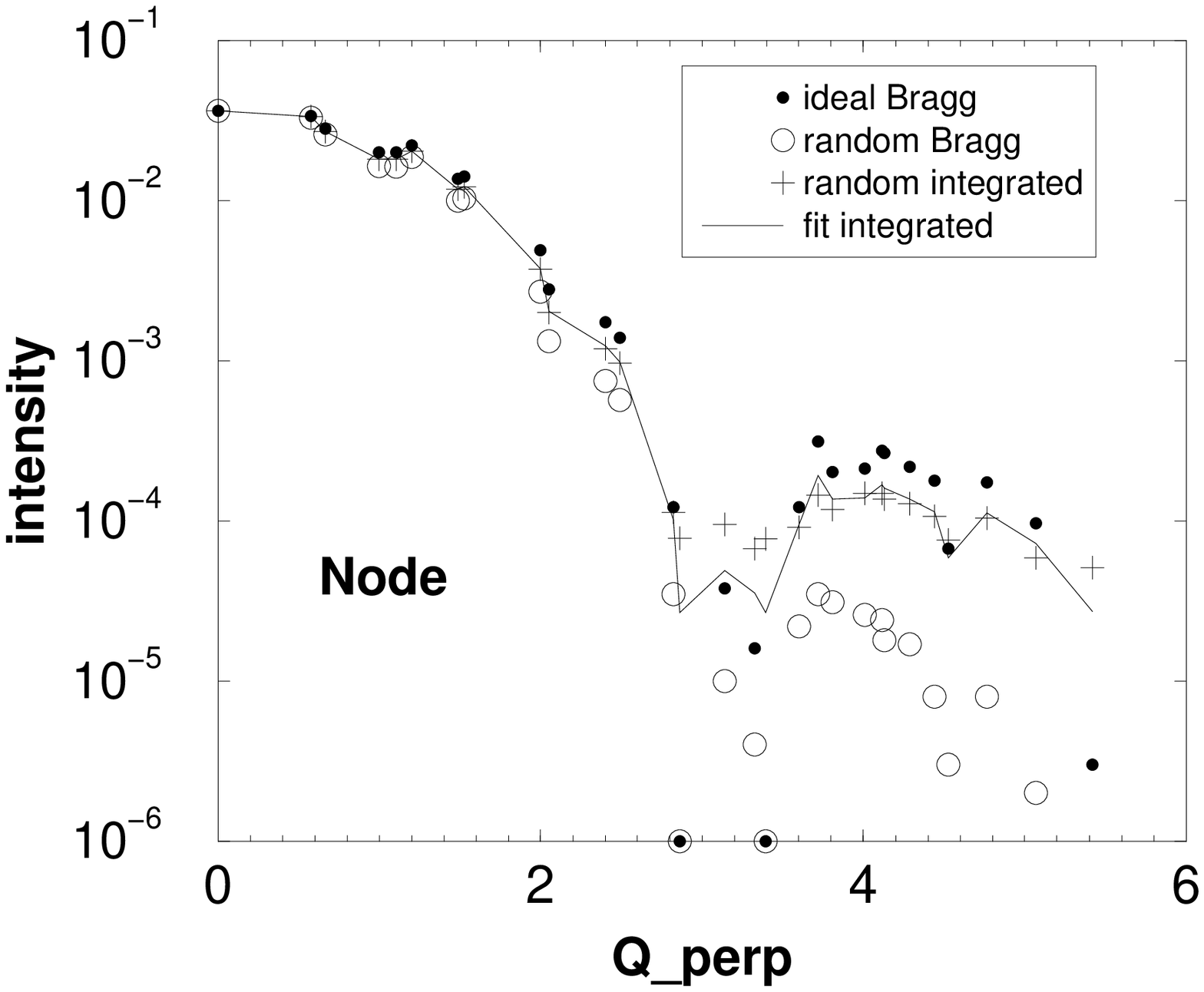}
   \epsfxsize=8.4 cm \epsfbox{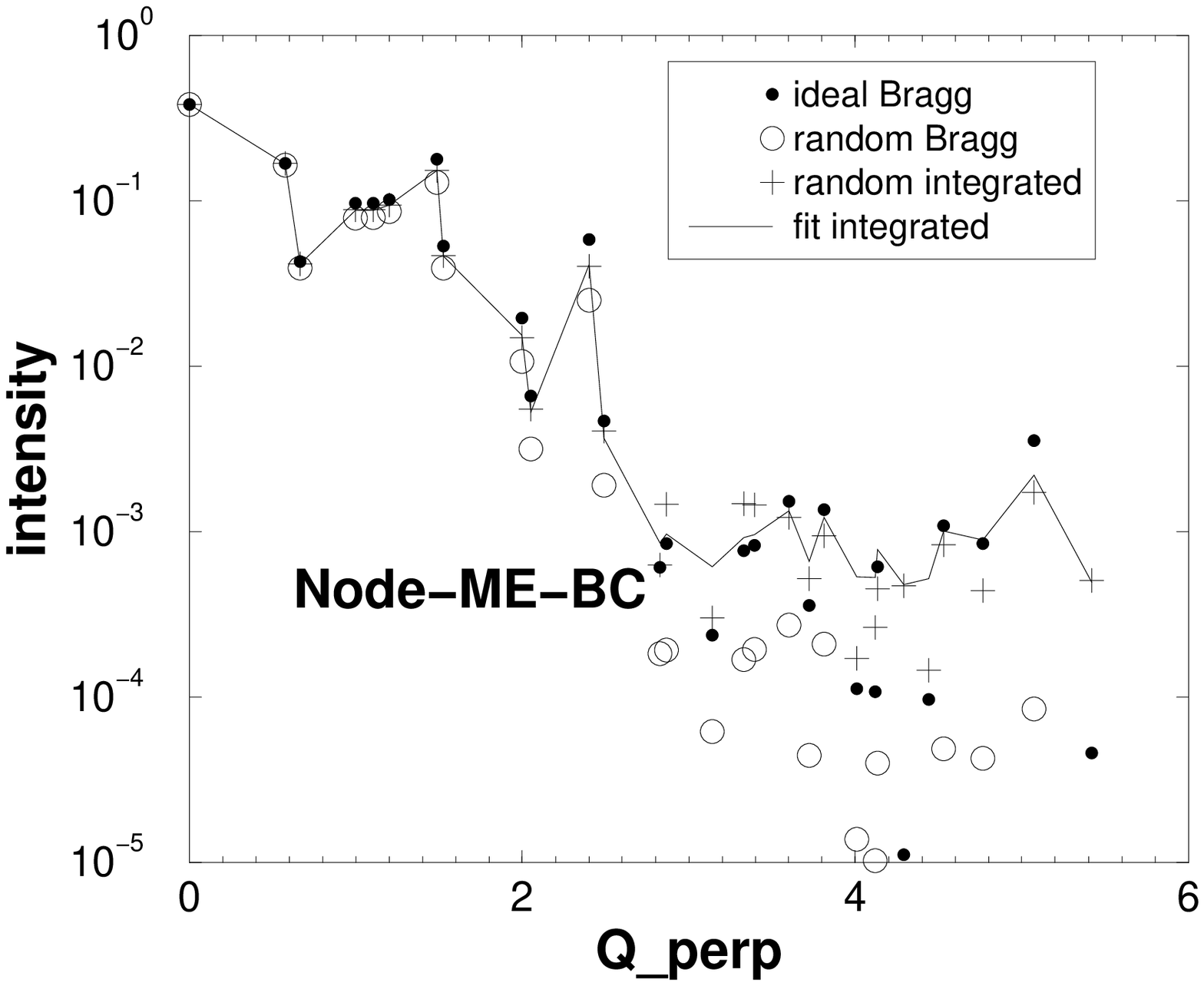} }
\caption
{Diffraction intensities from simulated ideal and random tilings,
as a function of perp wavevector.
Filled circles: Bragg peak intensities $I_{B,\id}$ 
from ideal 3D Penrose tiling (3DPT) of rhombohedra.
Empty circles: Bragg intensities $I_{B,\rand}$
from the random 3DPT. 
Crosses: total intensities $I_{I,\rand}$
from the random 3DPT, 
integrated over a sphere centered on 
the Bragg peak, to be compared with the solid line
representing eq.~\ref{eq:integrated}; see Sec.~\ref{sec:integrated}.
Plot (a) shows diffraction from the tiling nodes only, 
fitted by $c=0.51$, $D=3.0 \times 10^{-5}.$
Plot (b) is from a decoration in which node points have 
form factor $+1$,  while the midedge and ``body-center'' points 
(on axis of each prolate rhombohedron)
have form factor $-1$; this is fitted by
$c=0.47$ and $D=4.7 \times 10^{-4}$ in eq.~(\ref{eq:integrated}).
\rem{DATA: see mvic_rtdiff.sin1103.}
Note: in this and the next two figures, 
$|\GGperp|$ is called ``Q\_perp'' on the figure axis label.)}
\label{fig:simulations}
\end{minipage}
\end{figure}

\begin{figure}
\begin{minipage}{13cm}
\hbox {
\epsfxsize=8.4 cm \epsfbox{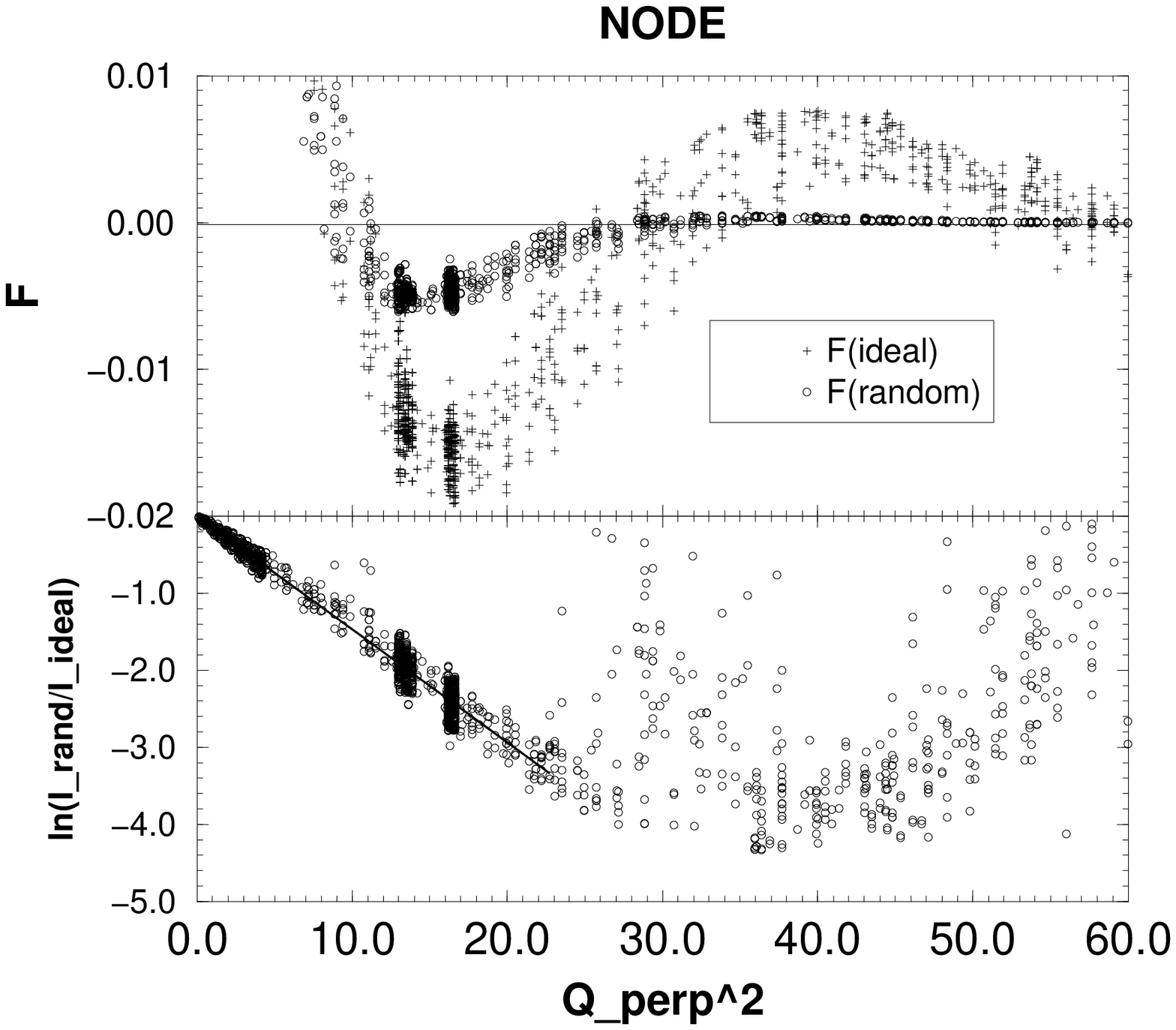}
\epsfxsize=8.4 cm \epsfbox{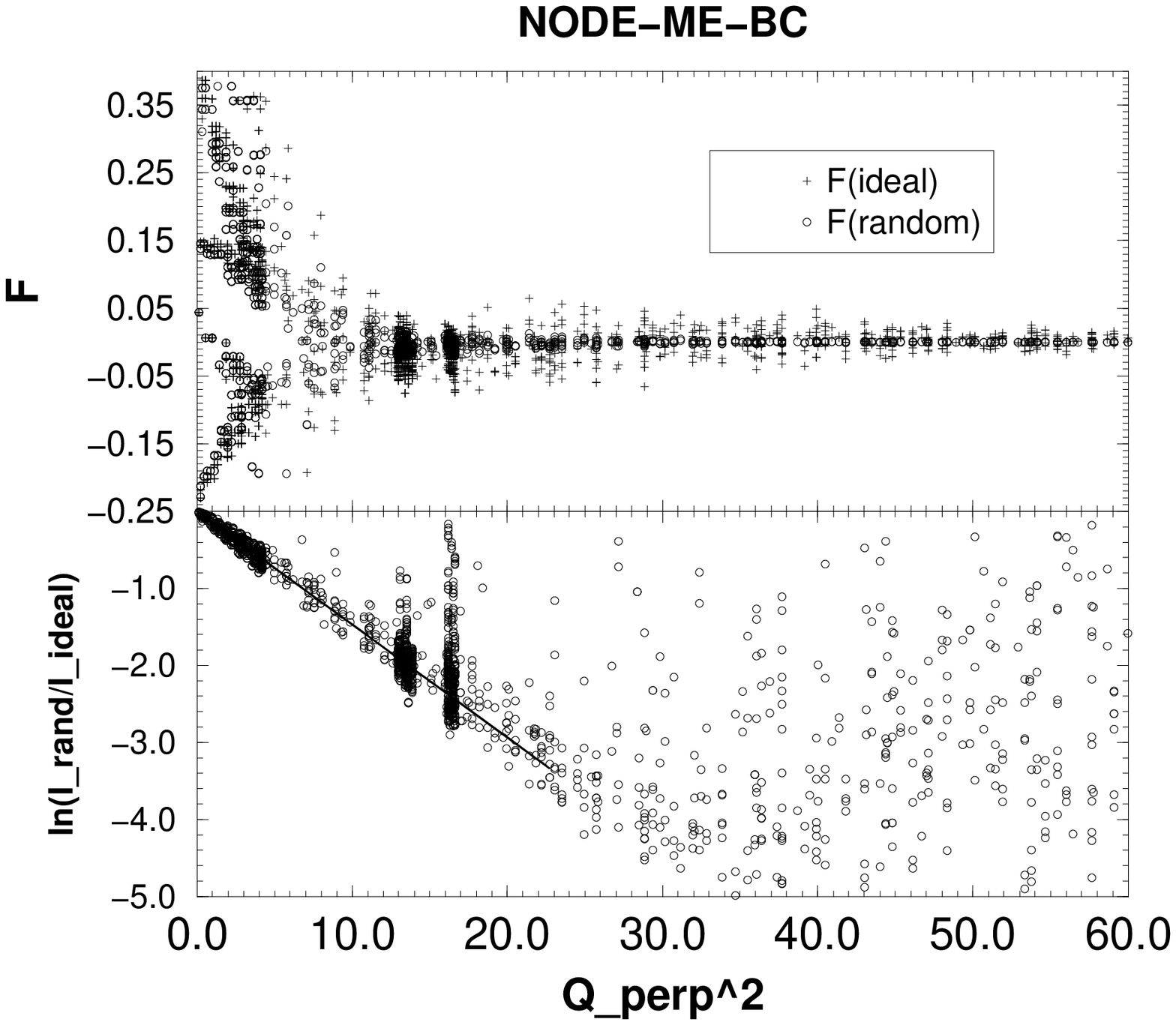} }
\caption
{Perp Debye-Waller factor in random tiling.
At top $|F(\GGperp)|$ is plotted for the ideal and the random 
tiling, showing the intensity reduction.
At bottom, the effective DW factor, $2 W_{\rm eff}$ from eq.~(\ref{eq:DWeff}),
is plotted against $|\GGperp|^2$ (labeled ``Q\_perp$^\wedge 2$'' in the figure);
its dependence is linear out to the second node of $F(\GGperp)$, showing the 
validity of the simple  formula (\ref{eq:DW}) out to surprisingly
large wavevectors.
Plot (a) shows diffraction from the tiling nodes only;
plot (b) is from a decoration in which node points have 
form factor $+1$,  while the midedge and ``body-center'' points 
(on the axis of each prolate rhombohedron)
have form factor $-1$.}
\label{fig:DW}
\end{minipage}
\end{figure}

Fig.~\ref{fig:simulations} shows the Bragg intensities plotted
against $|\GGperp|$ for selected wavevectors; the 
amplitudes (same data, but with signs) are plotted in 
the upper panels of Fig.~\ref{fig:DW}.
We can define the ratio
    \begin{equation}
         2 W_{\rm eff} \equiv - \ln \left[
         I^{\rm random}(\GGpar)/ I^{\rm ideal}(\GGpar) \right]
    \label{eq:DWeff}
    \end{equation}
which is plotted against $|\GGperp|^2$ in Fig.~\ref{fig:DW}(a).
The straight-line fit confirms (\ref{eq:DW}) with 
    $\sigma^2 = 0.145$. 
On the other hand, the variance $\langle |\hperp|^2 \rangle$
has been (easily) measured for both random and ideal tilings:
inserting these numerical values into Eq.~(\ref{eq:perpvardiff})
gives $3\sigma^2 (1.670)-(1.236) = 3(0.148) $, in perfect agreement.
The only caution to be mentioned is that the deviations which 
appear small because of the logarithmic scale of Fig.~\ref{fig:DW},
may in fact be far greater than the measurement errors of the 
intensities.  By the time the factorization approximation breaks
down completely, at $|\GGperp|^2 \approx 25$, the random-tiling
Bragg peaks (see top panel) are too small to measure in a real experiment. 


Do the Bragg amplitudes
behave differently when there are several orbits with 
different species of atoms, or when the
atoms do not sit on vertices?
To check this, we also calculated the diffraction from
a decorated model 
in which atoms with form factor $+1$
sit on all rhombohedron vertices and
another species with $-1$ sits 
on mid-edges and also 
on prolate rhombohedron axes at ideal points corresponding 
to body centers of six-dimensional cubes.~\footnote{
The positions are a simplification of 
Ref.~\onlinecite{Hen86}, but the assignment of species is different.
Negative form factors 
are realized by neutron diffraction for some isotopes.
We chose these form factors to make this diffraction pattern 
as different as possible from that of just the tile vertices.}
This places 54 120 atoms in the ``8/5'' size simulation cell.~ 

\begin{figure}
\begin{minipage}{8.66cm}
\hbox {
    \epsfxsize=8.4 cm \epsfbox{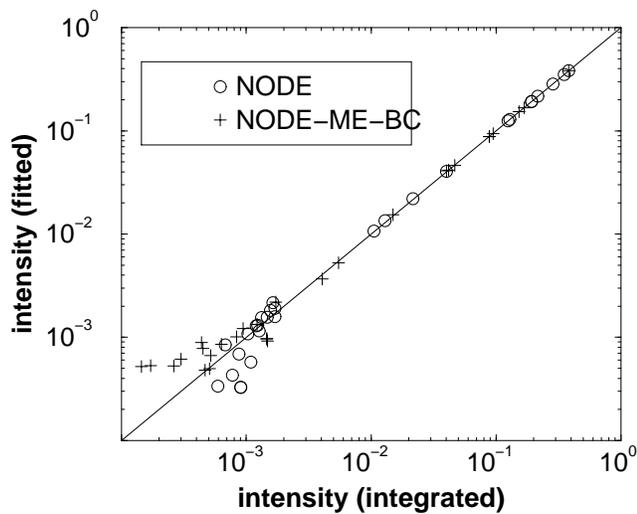}}
\caption
{The same data as in Fig.~\ref{fig:simulations}(a) are plotted here
(circles) as a comparison of the integrated intensity with 
the fit from eq.~(\ref{eq:integrated}), 
\MAREK{Check this.}
using the coefficients $c$ and $D$ mentioned in the
caption of Fig.~\ref{fig:simulations}(a). 
The crosses represent a similar comparison for the
decorated model shown in Figs.~\ref{fig:simulations}(b) and ~\ref{fig:DW}(b).}
\label{fig:integration}
\end{minipage}
\end{figure}

The answer -- appearing in Figures ~\ref{fig:simulations}, 
\ref{fig:DW},  and \ref{fig:integration} -- is that 
the decorated model is the same qualitatively --
for example, $\sigma^2$ is fit to $0.143$, essentially the
same value.  (The fit is just a little bit worse.)

\subsubsection {Correctness of the phase?}

A special concern was whether the random-tiling amplitude 
always has the same {\it sign} as the ideal-tiling.
Recall that, as a function of $|\GGperp|$, the ideal-tiling
amplitude oscillates and has zero crossings at certain wavevectors. 
The first zero-crossing of the random-tiling vertices
agrees precisely with that of the ideal-tiling vertices;
however, subsequent crossings at larger $|\GGperp|$ are noticeably different.
This indicates that the perp-space profile
of the atomic surface 
is {\it not} exactly given by (\ref{eq:convolve}).

\subsection{Integration over diffuse wings}
\label{sec:integrated}

Some of the intensity lost to diffuse scattering is
recovered when one integrates over the diffuse wings around
each Bragg peak; {\it some} such  integration is always performed
due to the resolution function.
We did another simulation to test
whether this, in some way, undoes the DW reduction.

We summed up the total intensity in a sphere centered on 
each Bragg point.  The sphere radius was 3 grid spacings of
the lattice of allowed wavevectors, 
so it contains 123 wavevectors.
\rem{2 pi/18.87 = 0.333}
(The spacing of our grid of discrete $\bf q$ points
is reciprocal to the cubic lattice constant 18.87 of the $8/5$ approximant.
i.e. $|\Delta {\bf q}| = 0.333$ in units of inverse 
rhombohedron edge. As is visible  in Fig.~\ref{fig:braggdiffuse}, 
the separation between noticeable Bragg peaks is 6  or 10
grid spacings, i.e. $\tau^{-1} \pi$ or $\pi$, so these spheres do not overlap.)
Due to computer time limitations, 
we only performed this computation for 
27 Bragg vectors, 
selected to be representative of the range of
$\Gperp$ values -- 
a very small fraction of all the usable Bragg vectors.
\rem{DATA for 27:  mvic_rtdiff.sin1103.}

Fig.~\ref{fig:simulations} also shows these integrated results (crosses).
Unexpectedly, the integration seems to restore the same fraction $c$
of the lost intensity of every peak 
{\it independent} of $\GGperp$ --
the same when $|\GGperp|$ is so small that 
the loss is invisible on the figure, and out to peaks where the ideal intensity is
down by $\sim 10^{-2}$ and the random Bragg intensity is
down a further $\sim 10^{-2}$ from the ideal. 
(Obviously, though, $c$ depends on the radius of our
integration sphere.)
The formula for this is 
   \begin{equation}
         I_{I,\rand} = c I_{B,\id} + (1-c) I_{B,\rand} + D , 
   \label{eq:integrated}
   \end{equation}
where $I_{B,\id}$ and $I_{B,\rand}$ are the Bragg intensities
for the ideal and random tilings, respectively, and
$I_{I,\rand}$ is the integrated intensity for the random tiling. 
To get a good fit, we had to include the constant $D$ 
which represents some diffuse intensity 
spread uniformly over reciprocal space.~\footnote{
Uniform diffuse intensity implies, of course,  a
component of the density that is completely uncorrelated 
from point to point.
We know no specific defect that
does this in our random tiling; the
alternation between the two ways of packing 
four rhombohedra into a rhombic dodecahedron 
might approximate this.}
Formula (\ref{eq:integrated}) is plotted for comparison in
Fig.~\ref{fig:simulations}; 
Fig.~\ref{fig:integration}
confirms more directly its remarkable success
(which is somewhat worse in the decorated model).

It is intriguing to speculate how a fit using the integrated
intensities would behave if
(\ref{eq:integrated}) were exactly correct for real experimental data, 
particularly if (due, say, to extinction problems) the strong peaks
at small $G_\perp$ were omitted from the fit. Then
$I_{B,\rand}$ could be neglected in (\ref{eq:integrated}) 
and $I_{I,\rand} \propto I_{B,\id}$, i.e., the ideal
structure would be reconstructed (apart from a
wrong factor of $c$ in the absolute normalization
of the density). 

Note that, inserting the phason Debye-Waller hypothesis 
of~(\ref{eq:DWreduced}) and (\ref{eq:DW}) into 
(\ref{eq:integrated}) implies
   \begin{equation}
         I_{I,\rand}/I_{B,\id} \approx c + (1-c)  
           \left(\exp {-\sigma^2 |\GGperp|^2}\right)
   \label{eq:DWfails}
   \end{equation}
Thus the {\it integrated} intensities are {\it not} expected to be fitted
by the phason Debye-Waller form
         $I_{I,\rand}/I_{B,\id} \approx \exp ({-\sigma_I^2 |\GGperp|^2})$.
However, if the $|\GGperp|$ values are so small
that $\exp({-\sigma^2 |\GGperp|^2}) \approx 1 - \sigma^2 |\GGperp|^2$, 
the fit will give an apparent $\sigma_I^2 \approx (1-c) \sigma^2$.
Indeed, a smaller value $\sigma^2$ was fitted by de Boissieu for 
low-resolution 
data than for high-resolution data in Ref.~\onlinecite{Bois94b}.

\subsubsection{Overall transfer to diffuse intensity}

A key issue in Sec.~\ref{sec:direct-veit}
was what fraction of intensity is transferred to 
diffuse scattering in a random tiling.
We investigated this  
using the same random rhombohedral tiling.
We summed the total diffuse intensity and the total intensity
for all wavevectors lying in the 2-fold plane.
The fraction of the total intensity that was ``diffuse''
was 0.145 in the 3/2 approximant and
0.095 in the 8/5 approximant: 
the 3/2 approximant shows a strong finite-size effect.
This fraction is artificially low in the 2-fold plane 
which contains many reciprocal lattice vectors with
small $\GGperp$. 
The fraction of {\it all} intensity which was diffuse was
$0.388$ in the 3/2 approximant;  it was too much to 
compute in the 8/5 approximant, so we can merely guess
it is $\sim 0.25$ (based on the trend in the 2-fold plane). 

\subsubsection {Digression on diffuse wings}

A great deal more can be learned if one measures the
detailed shape of the diffuse wings at high resolution, 
instead of just integrating over them.
Diffuse scattering, unlike Bragg scattering, 
depends on the spatial correlations of disorder.

Elastic theory governs the long-wavelength fluctuations 
of $\h(\r)$ in random-tiling-like quasicrystals;
energy-stabilized models
seem {\it not} to predict a gradient-squared elasticity
\cite{Lub88}.
Also, the shape of the diffuse wings of Bragg peaks is a 
key prediction of the elastic theory
\cite{diffuse91,JarNel88,Moriet91}. 
Thus, the match between experiment and elastic theory in
$i$-AlPdMn and in $i$-AlCuFe,~\cite{Bois95}
is evidence that these quasicrystals have a random-tiling nature.
(Some of us recently commented on these experiments and 
calculated some elastic constants, in 
\onlinecite{Mih99} and \onlinecite{Zhu99}.)

However, such analysis of diffuse scattering is outside 
the scope of this paper, which is 
crystallographic investigations based purely on {\it Bragg}
intensities of many peaks, aimed at discovering the microscopic
atomic arrangements.

\subsubsection {Global diffuse scattering?}

An unresolved question, and of relevance 
to background subtraction, is the diffuse background
far away from Bragg peaks.  It has been suggested by
Ref.~\onlinecite{Cap99} that one can account for
all of this
by summing over the diffuse wings of
{\it all} Bragg peaks, as approximated by 
elastic theory~\cite{diffuse91,JarNel88}), 
even from those that are farther away in reciprocal space 
than the typical separation of strong Bragg peaks.  
This is surprising since the elastic theory is 
a long-wavelength, coarse-grained theory that
ought to break down at the corresponding 
real-space distances (of the order of a tile edge).

In any case, this formula cannot be integrated over all
of reciprocal space, since even a single diffuse wing has
the schematic form $\int d^3 \qq (1/K|\qq|^2)$. 
It is of interest to write 
\newcommand{\Qmax}{{Q_{\rm max}}}
   \begin{equation}
   \langle |\hperp|^2\rangle = (2 \pi)^{-3} \int_0 ^ {\Qmax} d^3 \qq
       \frac{3{\tilde e}(K_2/K_1)}{K_1 |\qq|^2}  
   \label{eq:elasticcutoff}
   \end{equation}
where the 3 comes from taking the trace of a 
$3\times 3$ matrix, and ${\tilde e}(K_2/K_1)$ is a dimensionless factor
which averages out the angular dependences.
For $K_2/K_1 \ll 1$, it can be shown  that
${\tilde e}  = 1 + (8/9)(K_2/K_1)^2 + \ldots$, while it
diverges for $K_2/K_1 \to 3/4$ (the usual modulation
instability~\cite{diffuse91}).
Noting that $(K_1,K_2) =(0.8, 0.5)$ for
the rhombohedron tiling~\cite{Tang90,Shaw91} and
$\tilde e \approx 2$, we obtain $\Qmax \approx 3.5$
which is comparable to $\pi$ (the spacing between 
rather strong Bragg peaks).

\section {Review of tile decoration models}
\label{sec:review}

Before going on to the ``unified'' approach, we will need 
some more terms and concepts -- associated with 
tilings and their decorations (in real space)
-- to be used in the two sections
following.
To start off, 
we imagine that an ensemble of certain ``low-energy''
atomic configurations adequately represents
the states of the quasicrystal at temperatures below $1000 \rm K$. 
Then we insist that every ``low-energy'' 
configuration must correspond to a unique tiling. 
(Hence we sometimes say ``tiling'' as a shorthand for
``atomic configuration represented by the tiling,''
{\it e.g.}, when we speak of ``the tiling's energy.'')

\subsection {Decoration}

The mapping from the tiling to
an atomic configuration is called a {\it decoration}, since
each tile of a given class has atoms placed in
the same sites on the tile.  
If necessary, this placement is
allowed to have {\it context dependence} on the neighboring tiles.
But if the site is {\it sometimes} free to 
receive either of two species, depending on neighboring tiles, 
it may be preferable to consider this tile as having two
``flavors'' each with a unique decoration.
Sometimes atoms are assigned as decorations of other ``tile-objects''
such as vertices, edges, or faces,  
in order to reduce the number of site classes and numerical
parameters (see Sec.~II A of Ref.~\onlinecite{Mih96a}).
In all this, the principle is that all the constraints and
correlations are ascribed to the tiling level, so that 
the statistics of the tilings contain implicitly all the
statistics of the atom configurations.~\footnote{
Decoration models have been formulated such that
decorations of different tiles sometimes produce two atoms in the
same position, the rule being that these get replaced by one atom.
We do not consider such decorations, because
in analyzing or optimizing sums over the structure, 
giving either diffraction amplitudes or total energies, 
it is awkward when one cannot assign each atom to a well-defined
tile-object that it decorates.}

In random tilings, a rearrangement which turns one valid tile 
configuration to another, while affecting only a small cluster
of tiles, is called a {\it tile flip}. 
For a good decoration rule,  such that all tilings
in the ensemble have similar energies, there is a 
substantial agreement between the atomic configurations on those
tiles before and after the tile flip. 
(If this weren't so, one of the two configurations 
would have a much higher energy due to interactions with atoms 
in neighboring unflipped tiles.) 
Often, only a couple of atoms change positions in a tile flip.

\subsection{Supertiles}

The close overlap between the ``before-flip''
and ``after-flip'' atom configurations
often gives us freedom to change the size of tile while
only slightly changing the structure model.
When switching from large tiles to smaller tiles, 
the price of using small tiles is 
some simplifications in the decoration rule, 
and/or additional constraints among the tiles;
the price of using large tiles is having 
many sites on the tiles with independent parameters.
Replacing tiles by supertiles always entails a rebinding~\cite{Mih96a}
of their decorations.

The supertiling phenomenon is particularly prominent in decagonals
and is already familiar experimentally --
particularly in the various subphases of $d$(AlNiCo)~\cite{subphases}  --
and is also known as a cause of mis-indexing\cite{Lan94}.
The ``inflated'' tiles
in Penrose's tiling are a special case of supertile, but
most supertiles are not literal inflations.
Physically, starting from small tiles with a tile Hamiltonian, 
the small energy differences between tilings 
favor a sub-ensemble of degenerate lowest-energy configurations 
and often these may be represented
as tilings with bigger ``supertiles'', each decorated in a single
way with the original tiles;  a special case is
the interactions which favor a local pattern 
or ``cluster''~\cite{Jeong94}. 
Further remarks on supertilings are in Ref.~\onlinecite{Roth97}.

\subsection{Recipe to discover basic structure}
\label{sec:recipe}
\label{subsec:whichtiling}

\begin{table}[hbt!]
\label{tab:tilings}
\caption[]{Important tilings for realistic structure models.
Under ``Local?'', a ``yes'' means a {\it local update} exists; 
only such tilings are easy to simulate.
(We have sometimes called the
hexagon-boat-star tiling ``Two-level'', or
the rhombohedral tiling ``3DPT''; the latter
may also contain rhombic dodecahedra.)
Under ``symmetry'', ico, 10, and 12 are short for
icosahedral, decagonal, and dodecagonal, respectively.
References are indicated for tilings and decorations.}
\def\Strut{\large\strut}
\small
\begin{tabular*}{\linewidth}{lccl}
\hline
Tiling & Symm. & Local? & decorations  \Strut\\
\hline
rhombohedral  
\onlinecite{Tang90,Shaw91} &
ico & yes &
$i$(AlPdMn): \onlinecite{Els97} \\
canonical cells
\onlinecite{Hen91CCT,Mih93,New95} &
ico & no &
$i$(AlMn): \onlinecite{Mih96a,Mih96b} \\
binary 
\onlinecite{binary,Strand89} &
10  & yes &
$d$(AlCuCo): \onlinecite{Burkov} \\
hexagon-boat-star 
\onlinecite{Tang89} &
10 & yes &
$d$(AlMn): 
\onlinecite{LiKuo88} \\
& & & 
$d$(AlPdMn): \onlinecite{LiKuo88} \\
& & &
$d$(AlCuCo): \onlinecite{Coc97} \\
& & &
$d \rm (AlNiCo)$: \onlinecite{Wid00} \\
rectangle-triangle  
\onlinecite{Ox98} &
10 & no &
$d$(AlPdMn):  \onlinecite{Ox98} \\
& & &
$d \rm (AlNiCo)$ \onlinecite{Mih00}  \\
square-triangle 
\onlinecite{Ox93,sqtri} &
12 & no &
$dd$(VNiSi) \onlinecite{Kuo88} \\
& & &
$dd ({\rm Ta_{1.6}Te})$\onlinecite{ddTaSe}\\
\hline
\end{tabular*}
\end{table}

To proceed, one must first decide which tiling
geometry best represents the structure and 
its degrees of freedom.
Table ~\ref{tab:tilings} lists a menu of 
available random tilings which have been used  
in decoration models of real materials. 
(Some of the decorations were designed for a quasiperiodic
tiling, but are compatible with the random one.)
For the tilings marked ``no'' 
in column 3 of the table 
{\it any} simulation is just barely practical,
since each update move must involve an 
{\it indefinite} number of tiles.~\cite{Mih99,Ox93,Ox98}.

The proper size of the tiles is not self-evident even if, 
say,  one already knows the exact
structure of a large approximant crystal. 
Consider the most extreme case, a structure that looks
just like the Penrose tiling.  That structure may be
broken into tiles at any level of inflation; furthermore, 
at each level, the tiles may be represented 
as Fat/Skinny rhombi, as Kites/Darts, as Hexagons/Boats/Stars, 
or in a couple of other ways. 
At which level can we break apart the tiles and put them
together differently, to build other approximants or random tilings?
Only a computation (or intuition) of the energies can tell us. 
\rem{It has recently been asserted that d(ZnMgHo) -- is it Ho? --
has no ``clusters''.  What is really meant is that the atom
positions themselves happen to form a Penrose tiling, but it is
likely that 
the energy cost of rearrangements at that scale are prohibitive.}

In the rest of this section, we outline a 
recipe to find the right tiling, and a coarse version of the
structure model, {\it before} using any diffraction data, 
from simulations of lattice gases~\cite{Wid00},
or of random tilings with small tiles,
when microscopic Hamiltonians are available.
There are four or five stages.

\subsubsection
{Total energy code}
Before starting, one must have a means
to compute a total energy for any plausible configuration of the
atoms forming the quasicrystal.
Normally, the best choice is an effective pair potential
fitted to, or derived from, ab-initio data~\cite{pair-pot}.
In principle one can use ab-initio energies directly~\cite{Hennig99x}, 
but these are limited to very small system sizes. 

\subsubsection {Lattice gas simulation}
Construct a list of (available) sites, forming a quasilattice.
The quasilattice constant should be 
smaller than the interatomic spacing, so that the 
occupied fraction of sites is fairly small compared to unity.
A Monte Carlo simulation is carried out (using the 
Hamiltonian described in Stage 1), in which 
atoms are allowed to hop as a lattice-gas among these discrete sites;
(usually) the temperature is gradually reduced to zero and
the final configuration is  examined by eye.~\cite{Coc98}.
This annealing must be repeated over and over, since the
system will end up in different final configurations.
(That is natural when there are many nearly degenerate states, 
separated by barriers.)


Although the states are not constrained to be tile decorations, 
it may turn out that the low-energy states can be represented as such.
Discovering this representation is an art, and there may be more than one
correct answer (in that different tiling/decoration combinations
might generate identical atomic structures.)
The set of possible local patterns found in the 
resulting  tilings
implicitly defines a packing or matching rule for the tilings.
Thus, the lattice gas simulation serves as a systematic
procedure to discover {\it both} the ``right'' tiling (with its rules)
and the ``right'' decoration (relating the tiling to atomic positions).

A hybrid random tiling-lattice gas simulation
significantly improves on the plain lattice-gas simulation.
The system has two kinds of degrees of freedom (both discrete).
The first kind is a random tiling (which does {\it not} need to be
the same as the random tiling we are trying to discover by
this technique.)  Each tile is decorated in a deterministic
way by candidate sites: these sites take the 
place of the quasilattice defined in the plain lattice-gas
technique. The second kind of degree of freedom is a lattice
gas on these sites.  The Monte Carlo simulation must allow
tile flips as well as atom hops.

A different kind of hybrid simulation is to use a random tiling
of rather small tiles, with a deterministic decoration by atoms.
In effect, this is similar to a lattice-gas simulation, in that
(with around one atom per tile), practically every topological
arrangement of atoms is represented by some tiling.

\subsubsection
{Simulations of decorated random tilings}
At this stage, the degrees of freedom are tiles, and each tiling
is decorated deterministically with atoms.
A Monte Carlo annealing is perfomed in which 
all tilings are allowed, but are weighted as usual
by the interaction energy of their atoms.
Different variations of the decoration rule are
tried out, in an attempt to find the rule which gives
final states with the lowest energy.
This stage, like stage 2, 
refines {\it discrete} degrees of freedom. 
Also like stage 2, it finds both a decoration rule and
tile packing rules, but in  a less tentative fashion.

Having carried out stage 3 in small approximant crystals, 
one then repeats it in larger approximant crystals.
(Data from the small approximants could be misleading as
there are many local patterns of tiles which can fit in 
a large approximant or an infinite tiling, but not in 
a small approximant.)

\subsubsection
{Optimization of continuous parameters}
In this final stage, the decoration is fixed, but 
decoration parameters such as 
the atomic coordinates on the tiles are adjusted so
as to minimize the total energy.
One should still decorate more than one of the final 
annealed tilings from stage 3.

After stage 4, the decoration and its parameters are
finally fixed; the only degree of freedom is tile
configurations.  

\subsubsection 
{Construction of tile Hamiltonian}
The energy can be calculated 
for each possible tile configuration
by decorating it with atoms
and using the interatomic pair potentials, but this
is time-consuming for a large-size approximant.
It may be possible to find, and use, a ``tile-Hamiltonian''
which is an explicit (and simple) function of the
tiles or pairs of neighboring tiles, which accurately
mimics the pair-potential energy~\cite{Mih96a,Mih96b}.
This stage could be skipped.

\section {Unified fit from simulations} 
\label{sec:unifit}

The ``unified'' approach is the second of the two paths 
for structure determination described in this article.
It may be applied after one has a rough or moderately
detailed picture of the atomic structure and of the tile
degrees of freedom (from related approximants, for example). 
As noted in the first path, the procedure for fitting the averaged
density,  after the phase problem is solved, involves fits
very similar to those required 
by the ``unified'' method, and thus does not deserve a
separate discussion.

\subsection {Random tiling simulation}

An honest approach to a random tiling -- more 
precise than the
``factorization'' approximation of Sec.~\ref{sec:factorization}
-- requires a simulation, which 
can be used in several different modes.
Mode (i) would be to assume {\it a priori}
a particular random tiling ensemble (that means a 
fixed list of configurations sampled from a simulation);
only the decoration parameters are varied. 
\rem {In this case, 
the labor is essentially the same as for a decoration approach to a
{\it non-random} tiling.
If we used rational approximants rather than
perp-space analytic integration to carry out the Fourier transform).}
Mode (ii) would be to hold the decoration parameters fixed 
and vary the ``tile Hamiltonian'' parameters to control
tile correlations, until an optimum fit is achieved.
In this case, every iteration requires a fully 
equilibrated Monte Carlo simulation of the random tiling;
the approximant cell size may be limited by time constraints
since we must do repeated Monte Carlo simulations. 
Mode (iii) would be a ``simulated annealing''
of the tiling: one performs a Monte Carlo simulation
in which the $R$-factor itself replaces the tile 
Hamiltonian in the Boltzmann factor.
As the temperature is reduced, only the tile configurations
which optimize the $R$-factor will be represented.

\subsubsection {Structure determination assisted by energy calculations}

We strongly believe in the need to combine diffraction and
energy inputs to structure determination, because they
contain complementary information.~\footnote{
The familiar invocation of ``steric constraints'' to rule out 
unphysically close atoms could be considered an informal 
application of energy information.}
For example, some rare atoms are impossible to pin down by diffraction, 
but very easy to decide on the basis of energies. 
Again, transition-metal atoms from the same row ordinarily
cannot be distinguished by X-ray diffration, but they
can be clearly assigned by the use of potentials\cite{Wid00}.

In fact, exactly the same strategies can be applied either to the energy 
values or to the diffraction data. 
For example, one could run through all the steps of the 
``Recipe'' in Subsec.~\ref{sec:recipe} minimizing, instead of the
total energy, the $R$-factor for the fit to a diffraction 
data set.
(In a diffraction fit, 
the parameters appearing in the Stage 3 description 
might include Debye-Waller factors or partial occupations.)
While it is desirable to run energy and diffraction fits
separately, the ``most realistic''
fit should 
do them in parallel, minimizing some linear combination
of the total energy and the $R$-factor (or the $\chi^2$)
quantifying the mismatch to the measured Bragg intensities.
(This would be similar in spirit to the method of
``least-squares with energy minimization'' in
ordinary crystallography~\cite{LSQEMin}.

\subsection{Technical issues in simulations for 
diffraction fitting}

Here we discuss how one might, in practice, do an iterative
calculation which combines (i) Monte Carlo
reshuffling of the random tiling;
(ii) computation of Fourier sums to obtain the structure factors;
and (iii) adjustment of parameters to optimize the fit.

\subsubsection {Simulation cell as an approximant}

A simulation, of course, can only be done in a finite system, 
normally with periodic boundary conditions. 
Periodicity forces a net background phason strain;
this is minimized when
the simulation cell has the size and shape of
the unit cell of a good rational approximant of the 
quasiperiodic tiling. 
To calculate the diffraction, 
we repeat this cell's configuration throughout space.
Hence reciprocal space has a very fine grid of Bragg peaks, 
but most of them have negligible intensity even for
an approximant of an ideal quasiperiodic  tiling.
Of the Bragg peaks identifiable in the ideal case, 
most are lost in the random-tiling case. (Their intensity is greatly
reduced by what is loosely called the perp-space Debye-Waller factor, 
while meanwhile a uniform diffuse background appears almost
everywhere in reciprocal space; the Bragg intensity is 
overwhelmed if it is several times smaller than the diffuse intensity.)

Each orbit of equivalent quasicrystal Bragg peaks breaks up into 
several orbits of inequivalent Bragg peaks of the approximant.
We should average their amplitudes, to reduce the systematic
error (due to phason strain) and the statistical error (from the finite
size or time of the random-tiling simulation).
Before doing such an averaging, one must be careful to shift the 
unit cell so that the different approximant amplitudes have the
same phase. 
If the approximant is big enough, the inequivalent amplitudes
should be nearly the same. 

The deviations from symmetry always grow with the perp-space 
reciprocal lattice vector $\GGperp$. 
Meanwhile, the intensities decrease with $\GGperp$.  
So the simulation cell
should preferably be big enough so that, with increasing
$\GGperp$,  the intensities become
unmeasurable (or lost in the diffuse scattering) before the
deviations from icosahedral symmetry become overwhelming.
But often the simulation cell must be small, for the 
technical reasons we turn to next.

\subsubsection {Structure factor sum}

In all cases, 
one must re-evaluate structure factors repeatedly --
in the case of mode (iii), at every {\it step} of the iteration.
Even if the cell has relatively few tiles, it still
has more atoms than any unit cell of a metal crystal.
(For example, for the canonical cell tiling, the smallest reasonable
approximant is the 3/2 cubic cell, which contains typically 32 tile
nodes (112 tiles), but some 2500 atoms.)
Summing the exponentials is relatively expensive in computer time,
and the Fast Fourier Transform appears to be useless since
the sites do not lie on a simple grid.
Thus, the simulation cell size may be limited by the diffraction sum.

One can save work by partially factorizing the sums.
Let the index $\mu$ label the type of tile (or  tile-object), 
and let the (many) allowed orientations of each tile-object 
be indexed by $\omega$.
Let $n_{\mu}$ be the number of site type on tile-object $\mu$, 
according to the decoration rule in use, 
and let $\nu$ be the index labeling  them; also 
let $M_{\mu \nu}$ be the multiplicity of sites of type $\mu \nu$
on tile-object $\mu$.
Finally, let $N_{\mu \omega}$ be the number of tile-objects of type $\mu$
and orientation $\omega$ in the tiling.
Then we can define the ``tile structure factors'' as
   \begin{equation}
      \Ftile_{\mu\omega}(\q) = \sum _ {j=1}^{N_{\mu\omega}}
       \exp[i\q\cdot \R_{\mu\omega j}] 
    \label{eq:Ftile}
   \end{equation}
where $\R_{\mu\omega j}$ is the reference point of the $j$-th
tile-object  of type $\mu$ in orientation $\omega$. 

Also, we can define the ``decoration structure factors'' as
   \begin{equation}
     \Fdeco_{\mu\nu\omega} = \sum _ {i=1}^{M_{\mu\nu}} f_{a(\mu\nu)}(\q) 
    \exp [ i\q\cdot \u_{\mu\omega \nu i} ]
    \label{eq:Fdeco}
   \end{equation}
Here $\u_{\mu\omega i}$ is the displacement of the decorating atom
from the reference point of that tile-object, 
$a(\mu\nu)$ is the atom species in orbit $\mu\nu$, 
and $f_a(\q)$ is the form factor for species $a$.
Note that $\Fdeco_{\mu\nu\omega}$ 
for different orientations $\omega$ 
is trivially obtained  from some reference
orientation simply by applying the corresponding rotation or reflection
to $\q$. This symmetry can reduce the labor, though, only if that
rotation or reflection takes one of the allowed $\q$ vectors to another one.

The total structure factor then takes the form
   \begin{equation}
      F(\q) = \sum _{\mu\omega} 
      \Ftile_{\mu\omega}(\q)   \Fdeco_{\mu\nu\omega}(\q).
   \end{equation}

In a mode (i) process, the first factor remains fixed in every term;
in mode (ii), the second factor remains fixed.
Furthermore, it is easy to compute the {\it change}  in $\{ \Ftile_{\mu\omega} \}$
due to each Monte Carlo update move, and this usually affects relatively few of
the different $\mu\omega$ types.
Similarly, if we change a parameter in the decoration rule, 
we only need to update $\Fdeco_{\mu\nu\omega}$ for the orbit 
$\mu\nu$ which that parameter applies to.

\section {Decagonals}
\label{sec:decagonal}

Decagonal materials present special problems. 
\footnote{
The same considerations 
would apply to the other stacked quasicrystals
(octagonal and dodecagonal).}
The greatest problem is the seductiveness of
{\it apparent} two-dimensionality, which makes it
so much easier to visualize the tile packings than 
in icosahedral cases.  
In fact, however, the stacking is crucial.
{\it Physically}, it must amplify interactions 
within the layers and thus make the structure 
more ordered than a two-dimensional model could be
(and perhaps more ordered than analogous icosahedral
models are).
In {\it imaging} of the structure, averaging
may produce highly symmetrical patterns even when these
are not really present in individual layers.

\subsection {Stacking randomness}

First of all, the equilibrium random tiling phase of a decagonal
{\it cannot} be modeled by a random 
two-dimensional tiling~\cite{Hen91ART,Bur91}.
The entropy, which is supposed to stabilize such a phase,
would be proportional to the two-dimensional extent of the
system and (in the thermodynamic limit) would be negligible compared
to the volume.
(A corollary of this is that simulations of decagonals
with simulation cells just one lattice constant thick in the
$c$ direction do not give a valid picture of the tiling 
randomness. However they {\it are} valid for discovering which
atom arrangements are energetically favored, and which tiling 
is appropriate to model that composition.)

Instead, one must permit stacking randomness~\footnote{
We used the term ``randomness'' rather than ``disorder'', 
which would connote deviation from a particular, ideal structure.
In the random tiling case, the ideal structure is inherently
random. The stacking randomness is that a layer disagrees with 
the adjacent layers, not with an ideal layer.
We do {\it not} have in mind stacking disorder of the
Hendricks-Teller kind,  e.g. adjacent layers of type A 
when the normal pattern alternates type A and type B layers.}
that is, the tiling configuration is similar but not identical 
from one layer to the next. 
The difference between adjacent layers is constrained by rules
exactly analogous to the packing rules that determine how tiles
may adjoin within a layer. Unfortunately, stacking randomness
seems to be the least understood aspect of decagonal structures, 
both experimentally~\cite{Ritsch96,Witt99} and theoretically.~\cite{Shin93}
Like everything else in decagonals, the stacking rules might
depend sensitively on composition.

\subsection{Images}

We repeat an old warning about high-resolution transmission
electron microscope (HRTEM) images.
When a sample is thick
enough that a typical vertical section crosses a stacking change, 
then the image becomes (roughly speaking, and ignoring the 
dynamical effects) 
a two-dimensional projection of the scattering density. 
That projection is always perfectly quasiperiodic, and can not
reveal the randomness of individual layers.\cite{Hen91ART,Shaw91}. 

For this reason, we implore the practitioners of HRTEM (and
other techniques that project along the direction of the electron beam, 
such as the high-angle annular dark-field method 
discussed elsewhere in this volume):
please try to determine the thickness of the crystal!
The perfection of the image can be evidence for the perfection
of the quasicrystal only to the extent that an upper bound 
is placed on the thickness.

\begin{figure}
\begin{minipage}{13cm}
\epsfxsize=13cm \epsfbox{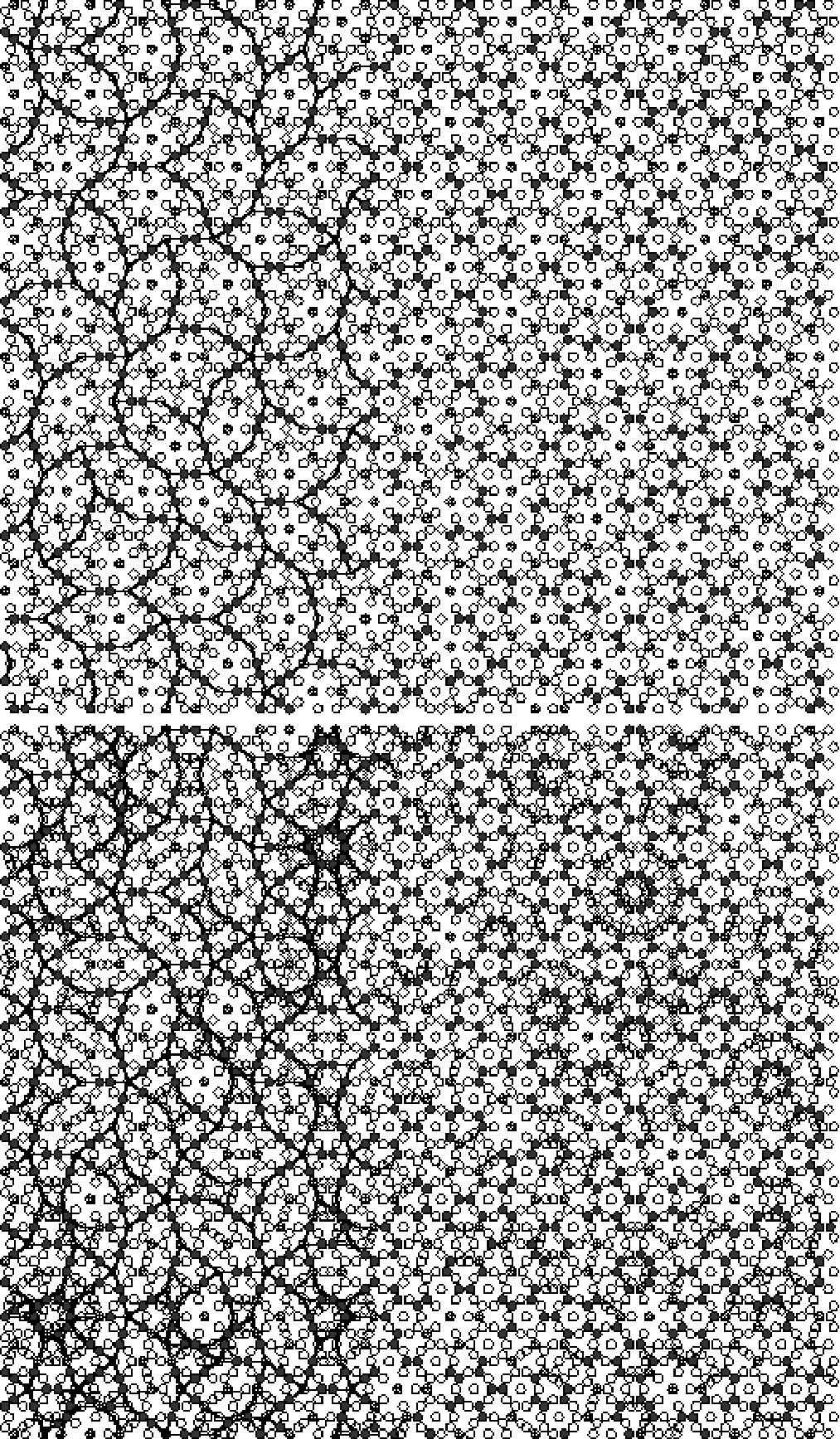}
\caption
{Top, a snapshot from one time step 
in a simulation of the (maximally) random Hexagon-Boat-Star tiling.
The tiling is decorated with atoms (Ni and Co shown filled), 
as in the $d$(AlNiCo) model of Ref.~\protect\onlinecite{Wid00}.
The tile edges are drawn in  only in the left portion.
Bottom, a superposition of nine successive time steps;
see how clusters of tenfold symmetry emerge that are
not present in the snapshot.
This is a poor man's imitation of a superposition of
different layers, as viewed in an HRTEM image.}
\label{fig:declayers}
\end{minipage}
\end{figure}

Multi-layer simulations of a realistic model of $i$(AlNiCo)
are feasible~\cite{Wid00} but not yet realized.
We propose that a superposition of time steps,
as shown in Fig.~\ref{fig:declayers}, is a good ersatz for
a superposition of layers.
The reason is that the 2D tilings in 
adjacent layers should differ by elementary tile
flips in isolated places, and the same thing us true for
configurations of time-evolving 2D tilings separated by small steps.
(The main qualitative difference is that the superposition of layers
is {\it less} random, at long wavelengths, than that of time steps, 
as discussed in Sec.~6.6 of Ref.~\onlinecite{Hen91ART}.)

The point that Fig.~\ref{fig:declayers} is intended to illustrate
is that clusters of (near) 10-fold symmetry emerge from the
projection which are {\it not} present in any layer.
Furthermore, to accomplish this, one does not need drastic 
differences from one layer to the next.
This is intended to be a caution about structure models
derived from electron microscopy, by assuming that the
image can be interpreted as a two-dimensional structure
(made of two layers).  Even more so, it is a caution
that the symmetrical ring motifs, which are so striking visually 
in the images, do not necessarily correspond to an atomic motif.

\subsection {Supertilings}

A final caution about interpreting HRTEM patterns will
be obvious to most of the microscopists, but perhaps
not to others.
{\it Supertiles} (see Sec.~\ref{sec:review})
are much commoner in decagonal than in icosahedral phases.~\footnote{
One reason may be that the natural inflation scale
for a decagonal 
is only $\tau=(\sqrt{5}+1)/2$ compared to $\tau^3$ for a
simple icosahedral, which as mentioned in Sec.~\ref{sec:perpdist}, 
is often appropriate in models of face-centered icosahedral structures.}
\rem {ANOTHER REASON: Also, if one identifies
``clusters'' to be nodes of the supertiling, the directions
of possible linkages between these clusters 
are always separated by $36^\circ$ in the decagonal case, just
like the original tile directions; in the icosahedral case, 
linkages in 2-fold, 3-fold, and 5-fold directions all have different
geometries.}
Quite typically -- in analyzing experimental images or
simulated configurations~\cite{Wid00,Coc99} --
the structure is manifestly a packing of certain small tiles (or clusters).
Yet they are highly constrained, so one suspects that a description in
terms of a supertiling would be more economical.  But which one is correct?
From initial data, perhaps, one can conjecture that certain 
configurations are forbidden, and we can show this 
implies that the allowed configurations are exactly 
decorations of a certain simple supertiling.
Yet we then notice (in larger simulation cells)
one or two exceptions, indicating that the real rule is more complicated. 
(Although wrong, the simple supertiling is not useless 
-- it might be quite adequate for a fit of the available diffraction data.) 

This observation is intended as a second caution about 
reducing HRTEM images to tilings by placing nodes
at the centers of pentagonal or decagonal rings of spots.
In the actual images, there is often a continuum of patterns
between (say) symmetrical rings and distorted rings: 
one must almost arbitrarily include some nodes and exclude others.
Small differences in this rule, or errors in its application, 
may induce major changes in the rules of the inferred tiling, 
in the sense that a previously forbidden local pattern may
start to appear, or vice versa.

\section {Conclusion}

All steps of the procedure for fitting a random-tiling
structure are now available in principle, but they
have not yet been put into practice. 
Even though a new kind of ``direct method'' may determine
the phases of Bragg amplitudes (Sec.~\ref{sec:direct-veit});
this merely produces the ensemble-averaged 
scattering density;
there remains the harder task of modeling the
random-tiling ensemble which has been averaged.
In accomplishing this, it seems hard to disentangle energy modeling
from structure modeling. 
Indeed we suspect that the best
structure fits of the future will combine total energy calculations
and diffraction refinements in some fashion. 

In principle, a simple perp-space Debye-Waller factor should 
{\it not} suffice to model the relation between 
the ideal and random tilings, but in practice this seemed to
work rather well for the rhombohedron tiling
(Sec.~\ref{sec:factorization}).
More study is needed: it would be interesting, for example,
to apply the ``minimum-charge'' method of Sec.~\ref{sec:direct-veit}
to the Monte Carlo ``data'' of Sec.~\ref{sec:factorization}.
And if one is given only the random-tiling data, what sort of
ideal tiling will be reconstructed from it if we adopt
the factorization approximation and divide out the best-fit
Debye-Waller factors?

Decagonal structures are deceptive.
They are easy to visualize using two-dimensional 
images, but the equilibrium  random ensemble is {\it never} 
represented by a two-dimensional tiling, 
It should be a priority -- both in experiment and in
total-energy modeling -- to understand the
stacking randomness in decagonals. 

At the very start of this paper, we demanded a model which contains
information beyond the averaged density: implicitly
that means information about correlations.
Yet we set ``rules of the game'' which restricted us to using
only Bragg data, discarding the diffuse intensity which is
in fact the Fourier transform of the correlation function.
So, an interesting question for the future is whether there
is some systematic way to incorporate {\it diffuse} data from
all of reciprocal space (not merely wings of Bragg peaks)
in a quantitative fit of parameters of a random tiling 
structure model.

\begin{acknowledgment}
We thank K. Brown and A. Avanesov for help in the
phase computations and for producing
figures \ref{fig:mackay} and \ref{fig:periodic}. 
We thank them, as well as R.~Hennig, M. de Boissieu, 
E.~Cockayne, M.~Widom, and M.~E.~J. Newman, 
for discussions, comments,  and collaborations. 
C.L.H. and M.M. were supported by DOE grant No. DE-FG02-89ER-45405. 
Computer facilities were provided by the Cornell Center for Materials
Research under NSF grant DMR-9632275. 
V.~E. was supported by NSF grant 9873214.
\end{acknowledgment}


\onecolumn

\eject

\eject

\end{document}